%% file: arxiv_sub.tex
\documentclass[a4paper, DIV14]{scrartcl}

\usepackage[utf8]{inputenc}
\usepackage[T1]{fontenc}
\usepackage[english]{babel}
\usepackage[dvipsnames]{xcolor}
\usepackage{authblk}
\usepackage{natbib}
\usepackage{graphicx}
\usepackage{soul}

\usepackage{xr}
\input{pkg_def.tex}

\author[1,2]{Paul F.V. Wiemann\thanks{corresponding author}}
\author[1]{Thomas Kneib}
\author[3]{Julien Hambuckers}
\affil[1]{Chair of Statistics, G\"{o}ttingen University, Germany}
\affil[2]{Chair of Computational Statistics, TU Dortmund University, Germany}
\affil[3]{HEC Li\`{e}ge, Department of Finance, University of Li\`{e}ge, Belgium}

\title{Using the Softplus Function to Construct Alternative Link Functions in Generalized Linear Models and Beyond}

\date{\today}

\begin{document}
\maketitle

\abstract{\input{abstract}}


\section{Introduction}

Response functions and their inverse, link functions, are central parts of many modern regression approaches.
They create a one-to-one mapping between predictors defined on the real line and the distributional parameters featuring restricted support (e.g., positivity restrictions).
The choice of the response function can affect model quality in two ways.
First, without a suitable response function the model may not fit the data given.
Second, the interpretability of covariate effects depends crucially on the response function.
For example, the identity function implies additive effects while the exponential function facilitates a multiplicative interpretation.
Complex response functions may even hinder any useful interpretation except the marginal interpretation, that is, assessing the effect of one covariate on the distributional parameter visually by keeping all but the covariate in question constant.

\begin{figure}[bth]
  \centering
  \includegraphics[width=.75\textwidth]{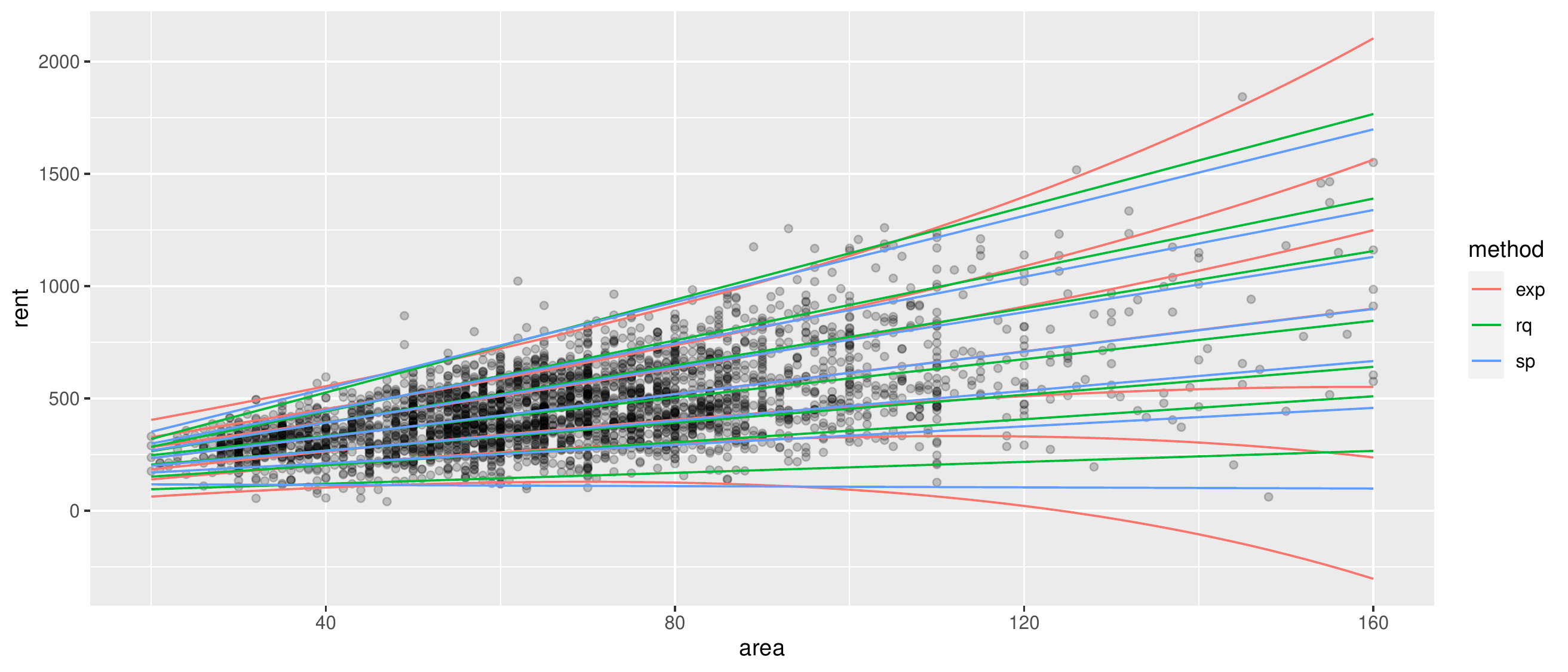}
  \caption{Scatter plot of the rents versus living area. Quantiles estimated for $q = 0.01, 0.1, 0.25,0.5,\protect\linebreak 0.75, 0.9, 0.99$  via quantile regression (qr) as well as location-scale regression with exponential (exp), and with softplus response function (sp) are shown as lines.}
  \label{fig:munich-rent}
\end{figure}

In this paper, we suggest to construct novel types of response functions for strictly positive parameters based on the softplus function.
These response functions can be used when an additive interpretation of effects is desired since they allow for a quasi-additive interpretation over a certain part of the range of predictor values. Consider as a motivating example heteroscedastic data from the Munich rent index plotted in Figure~\ref{fig:munich-rent}.
We relate rents in Euro of flats in Munich to the size of living area in square meters,
dealing with the heteroscedasticity by using a linear quantile regression model \citep{koenkerQuantileRegression2005} and two versions of location-scale models within the framework of generalized additive models for location, scale and shape \citep{rigbyGeneralizedAdditiveModels2005} assuming normally distributed responses with regression effects on both the mean and the standard deviation.
The location-scale models differ only in the response function employed for the scale parameter (viz., the exponential and softplus response function).
In every model, the predictors consist of an intercept and the linear effect of living area.
Figure~\ref{fig:munich-rent} shows estimated quantiles and, thereby, illustrates how the non-linearity of the exponential function translates to the estimated quantiles as a function of the living area.
In contrast, the softplus response function maintains the linear relationship in the predictor providing a parametric counterpart to the non-parametric quantile regression model.
The estimated quantiles are more similar to the results from the quantile regression model.
We explain the reasoning behind the quasi-additive interpretation in Section~\ref{sec:meth_softplus} and provide details on this application in Section~\ref{sec:munich-rents}.

For the choice of the response function, most researchers rely on default choices such as the logistic response function for parameters restricted to the unit interval (e.g., probabilities), or the exponential response function for strictly positive parameters.
In generalized linear models \citep[GLMs,][]{mccullaghGeneralizedLinearModels1989}, these defaults can often be justified by their characterization as natural link functions arising in the context of exponential families.
In other cases, the default response functions are chosen to entail specific modes of interpretation, e.g., multiplicative effects on odds in case of the logistic response function or multiplicative effects on the parameter of interest in case of the exponential response function.
The interpretability is also the reason why \cite{fahrmeirRegression2013} recommend to use the exponential response function for gamma distributed responses in the GLM framework instead of the canonical link function.
However, the use of the exponential response function implies a multiplicative model, contrasting with the assumption of additivity of effects often desired in statistics.
Moreover, domain-specific knowledge about the application can invalidate a multiplicative model entirely and e.g., suggest an additive model.
As shown above, an additive model for the standard deviation of a normally distributed response variable leads to additive effects on the quantiles of the response distribution, providing a parametric counterpart to non-parametric quantile regression specifications.

So far, researchers need to fall back to using no response function if the additivity of effects is desired.
This implies the aforementioned issues when the distribution parameter modelled must comply with a positivity restriction.
With this paper, we introduce the softplus response function to regression modeling to overcome these issues.
The softplus response function allows for a quasi-additive interpretation of regression effects for the majority of the relevant predictor space.
Nonetheless, it is a strictly increasing bijective function mapping the real values to its positive subset.
Therefore, it is an eligible response function for positively restricted distribution parameters and can be used instead of the exponential response function (guaranteeing a quasi-additive model) or the identity response function (avoiding the restriction of regression coefficients).

In addition to the quasi-additive interpretation, the softplus function enables the design of response functions with interesting properties.
It can be augmented with an additional parameter that yields further flexibility to model the data given, (ii) it avoids exponential growth which can be an issue under certain covariate combinations, and (iii) it enables the construction of an exponential-like function that avoids potential numerical overflow when evaluating it for large positive predictor values.

An alternative to pre-chosen response functions is to estimate the response function flexibly from the data.
The most well-known example for this approach is the single-index model introduced by \cite{ichimuraSemiparametricLeastSquares1993}.
The kernel-based single-index models share the disadvantage that the estimated response function is often to flexible.
To counter this characteristic, \citep{yuPenalizedSplineEstimation2002,yuPenalisedSplineEstimation2017} introduced penalization to single-index models.
Recently, \cite{spiegelGeneralizedAdditiveModels2019} presented an approach that combines the single-index models based on penalized splines with the flexibility of generalized additive models \citep{hastieGeneralizedAdditiveModels1986}.

One practical challenge when employing flexible link functions is the interpretation of the resulting model since restrictions have to be assigned to the regression predictor to render the response function estimate identifiable.
In contrast, simple, fixed response functions considerably facilitate interpretation.
Having easily interpretable effects may be the reason that for positively bounded parameters the exponential response function is still the most common approach.

Regardless of how well default choices can be justified in general, there is no a priori reason to assume that the defaults fit well on any given data set.
Therefore, the investigation of alternative response functions is a worthwhile and relevant endeavour.

The remainder of this paper is structured as follows: Section~\ref{sec:meth_softplus} introduces the softplus response function, justifies the quasi-additive interpretation and gives a guideline for its proper use.
Furthermore, Section~2 describes statistical inference when employing the softplus response function.
Section~\ref{sec:simulation} investigates the softplus response function in simulation studies.
The practical applicability of softplus-based regression specifications is demonstrated in Section~\ref{sec:applications}.
The final Section~\ref{sec:sumconc} summarizes our findings and discusses directions of future research.

\section{The Softplus Function}\label{sec:meth_softplus}

\begin{figure}[tb]\centering
\includegraphics[width=.75\textwidth]{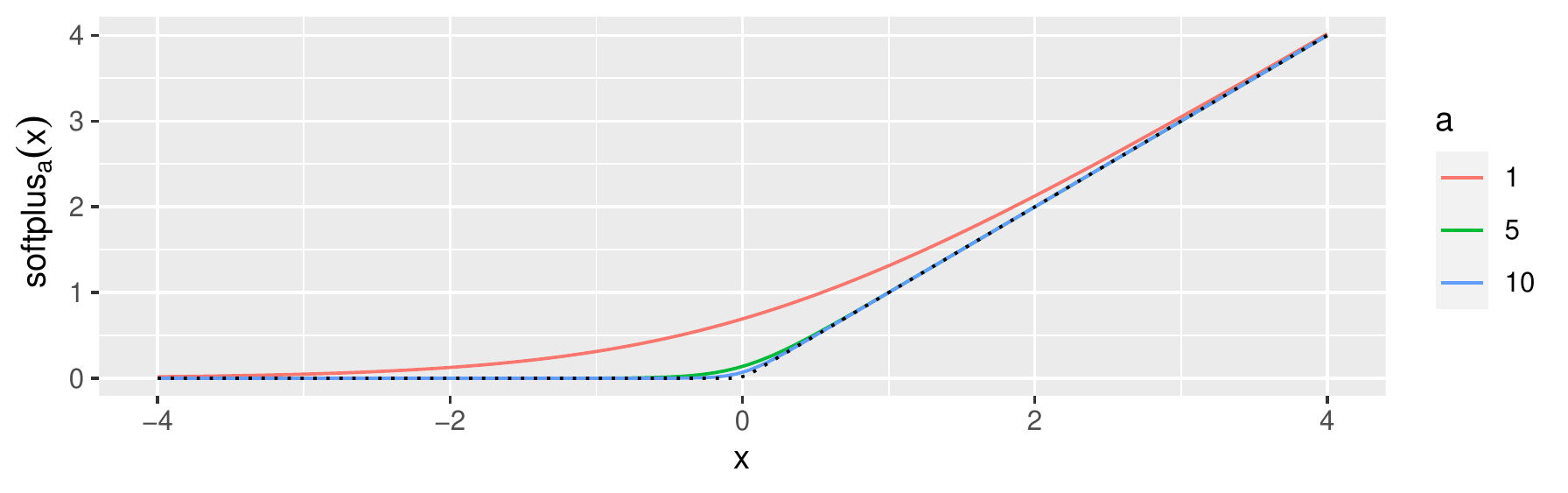}
\caption{\label{fig:softplus} Plot of the softplus function (left) for different values of softplus parameters $a$. The approximated rectifier function is shown as the dotted line.}
\end{figure}

So far, the softplus function \citep{dugasIncorporatingSecondOrderFunctional2001} is mainly used as a continuously differentiable approximation of the rectifier function (i.e., $\rect(x) = \max(0, x)$) in deep neural networks \citep{haozhengImprovingDeepNeural2015}.
The function maps the elements of $\mathbb{R}$ to elements of its positive subset $\mathbb{R}_+$. We use a generalized version which can be defined by the equation
\begin{align}
  \softplus_a(x) = \frac{\log\left(1 + \exp(ax)\right)}{a}\label{eq:def-softplus}
\end{align}
featuring an additional parameter $a > 0$.
Setting $a = 1$ reduces the function to its simple form.
Figure~\ref{fig:softplus} shows the softplus response function for different values of $a$.
Introducing the softplus parameter $a$ allows us to control the approximation error with respect to the rectifier function, as it can be shown that for every $\varepsilon > 0$, exists some $a > 0$ such that
\begin{align*}
0 < \operatorname{softplus}_a(x) - \max(0, x) \leq \log(2)/a < \varepsilon
\end{align*}
holds for all $x \in \mathbb{R}$.
The largest approximation error is at $x = 0$ as visually indicated by Figure~\ref{fig:softplus} and follows from the reformulation for numerical stability subsequently discussed (see Equation~\eqref{eq:sp_stable}).
Besides, one can observe in Figure~\ref{fig:softplus} that the softplus function follows the identity function very closely in the positive domain and rapidly approaches zero in the negative domain for $x\rightarrow-\infty$.
This behavior can be further accentuated by increasing the parameter $a$.
Therefore, the softplus parameter $a$ can be used to control how long the quasi-linear relationship should be maintained when approaching zero and consequently, how fast the boundary of the linked distribution parameter is approached.

The adoption of the softplus function as a response function features two major advantages:
\begin{itemize}
  \item It translates the additivity of effects on the predictor level to the parameter space for a majority of the relevant distribution parameter space while always guaranteeing the positivity of the distribution parameter.
  This is achieved by being quas-linear in its argument as long, as the predictor is large enough for a given value of $a$.
  \item The softplus function allows for a straightforward interpretation of the covariate effects. When the predictor value is large enough, the effects can be interpreted directly on the parameter.
    Consider the linear effect of an covariate $x$ with the corresponding regression coefficient $\beta$. An increase of $x$ by one unit is associated with an increase of $\beta$ in the predictor and when using the softplus function also an increase of almost $\beta$ in the distribution parameter or expressed as an formula ${\softplus(\beta(x + 1)) \approx \beta + \beta x}$.
\end{itemize}
Clearly, the quasi-additive interpretation is no longer valid once the argument of the softplus function is not within the approximately linear part of the softplus function which we define in detail in Section~\ref{sec:linear-part}.
However, by choosing a sufficiently large $a$, the linear part covers almost the entire positive domain.
In the negative domain and for a sufficiently large $a$, a small change of the covariate usually does not cause a relevant change on the parameter value, since the softplus function outputs values very close to zero.
To ensure the validity of this interpretation, it is necessary to check the range of values of the linear predictor for the observations in the data set.
Most of them should be located within the linear part of the softplus function.

The quasi-additive interpretation is in contrast to the usual multiplicative interpretation for positively constrained parameters that arises from the use of the exponential response function.
For example, $\exp(\beta (x + 1)) = \exp(\beta)\exp(\beta x)$ leads to the interpretation that a change of one unit in the covariate is associated with a multiplicative change of $\exp(\beta)$ units on the parameter.

In contrast to large values for $a$ which enable the quasi-additive interpretation, with the choice of sufficiently small values for $a$, the softplus function resembles the exponential function with a scaled and shifted argument.
This becomes more obvious when taking into account that $\log(x + 1)$ is almost linear in $x$, for $|x| \ll 1$, and thus $\log(1 + \exp(ax)) / a \approx \exp(ax - \log(a))$ for ${\exp(ax) \ll 1}$.
Consequently, the choice of the softplus parameter $a$ allows to continuously vary between an identity-like response function (for $a \rightarrow \infty$) and the exponential response function (with scaled and shifted argument for $a \rightarrow 0$).
This very property facilitates the construction of another response function approximating the exponential function for small arguments but with a limiting gradient (see Section~2 in the Supplementary Material).
This function can be used when an exponential-like response function is desired, but unbounded growth is an (e.g., numerical) issue.

\subsection{Numerical Stability of the Softplus Function}\label{sec:numerical}

A naive implementation of the softplus function derived from Equation~\eqref{eq:def-softplus} can easily lead to numerical issues.
The value of the exponential function for relatively small inputs is infinity on common computer hardware.
To give some intuition, according to the IEEE~754 standard \citep{zuras2008ieee} the largest 32-bit and 64-bit floating point numbers are roughly $3.4028 \cdot 10^{38}$ and $1.7977 \cdot 10^{308}$, respectively.
Consequently, calculating $\exp(89)$ and $\exp(710)$, respectively, yields infinity.

This is of special concern for the implementation of softplus function since the argument to the softplus function is multiplied with the softplus parameter $a$ before the exponential function is applied.
Consider a Poisson regression model with softplus response function and $a=10$.
The predictor $\eta = 9$, targeting an expected value of $9$, would already yield infinity on a 32-bit system using the naive implementation although the correct result is between $9$ and $9 + 10^{-40}$.
Albeit 64-bit CPUs are common nowadays, one should still consider 32-bit floating point arithmetic since it is often used in high-performance computing or when the computation is carried out on graphical processing units (GPUs) or tensor processing units (TPUs).

Despite the difficulties described, the softplus function becomes numerically stable by using the equality
\begin{align}
  \softplus_a(x) &= \max(0, x) + \frac{\log(1 + \exp(-|ax|))}{a}\label{eq:sp_stable}
\end{align}
in conjunction with the \texttt{log1p} procedure.
\texttt{log1p} evaluates $\log(1 + x)$ very precisely even for $|x| \ll 1$ \citep[p. 68]{abramowitzHandbookMathematicalFunctions1972} and is available in most programming languages.
In this formulation, the exponential function must be evaluated only for arguments less than 0 which can be done accurately.
Besides numerical stability, Equation~\eqref{eq:sp_stable} also implies that the softplus function has its largest approximation error with respect to rectifier function at $x=0$ with $\log(2)/a$.

The correctness of the numerical stable formulation is easily verified by expressing the softplus function in terms of the log-sum-exp (LSE) function and exploiting its translation property.
The LSE function takes $l$ real valued arguments $x_1, \dots, x_l$. Its value is given by
\begin{align*}
  \operatorname{LSE}(x_1, \dotsc, x_l) = \log\left(\sum_{i=1}^l \exp(x_i)\right).
\end{align*}
The translational property \citep{nielsenGuaranteedBoundsInformationTheoretic2016} states that for $c \in \mathbb{R}$
\begin{align*}
  \operatorname{LSE}(x_1, \dotsc, x_l) = c + \log\left(\sum_{i=1}^l \exp(x_i - c)\right)
\end{align*}
holds.
Consequently, we have
\begin{align*}
  \softplus_a(x) &= \frac{\log(1 + \exp(ax))}{a} = \frac{\operatorname{LSE}(0, ax)}{a}\\
    &= \max(0, x) + \frac{\log(\exp(0 - \max(0, ax)) + \exp(ax - \max(0, ax)))}{a}\\
    &= \max(0, x) + \frac{\log(1 + \exp(-|ax|))}{a}
\end{align*}
where the second line arises by setting $c = \max(0, ax)$ and the last line follows from the observation that $-|ax| = ax$ for $x < 0$.
We provide a numerical stable reformulation of the inverse of the softplus function in the Supplementary Material.

\subsection{Linear Part of the Softplus Function}\label{sec:linear-part}

Telling apart the section of the softplus function that is approximately linear is important when it comes to the interpretation of regression effects.
In its approximately linear part, the softplus function approximates the identity function and thus permits the quasi-additive interpretation.
A change in the predictor within this part can directly be interpreted as a change in the parameter.
However, since the softplus funtion is not linear over its whole domain we need to define under which conditions we consider this interpretation to be valid.

\begin{figure}[hbt]
  \centering
  \includegraphics[width=0.75\textwidth]{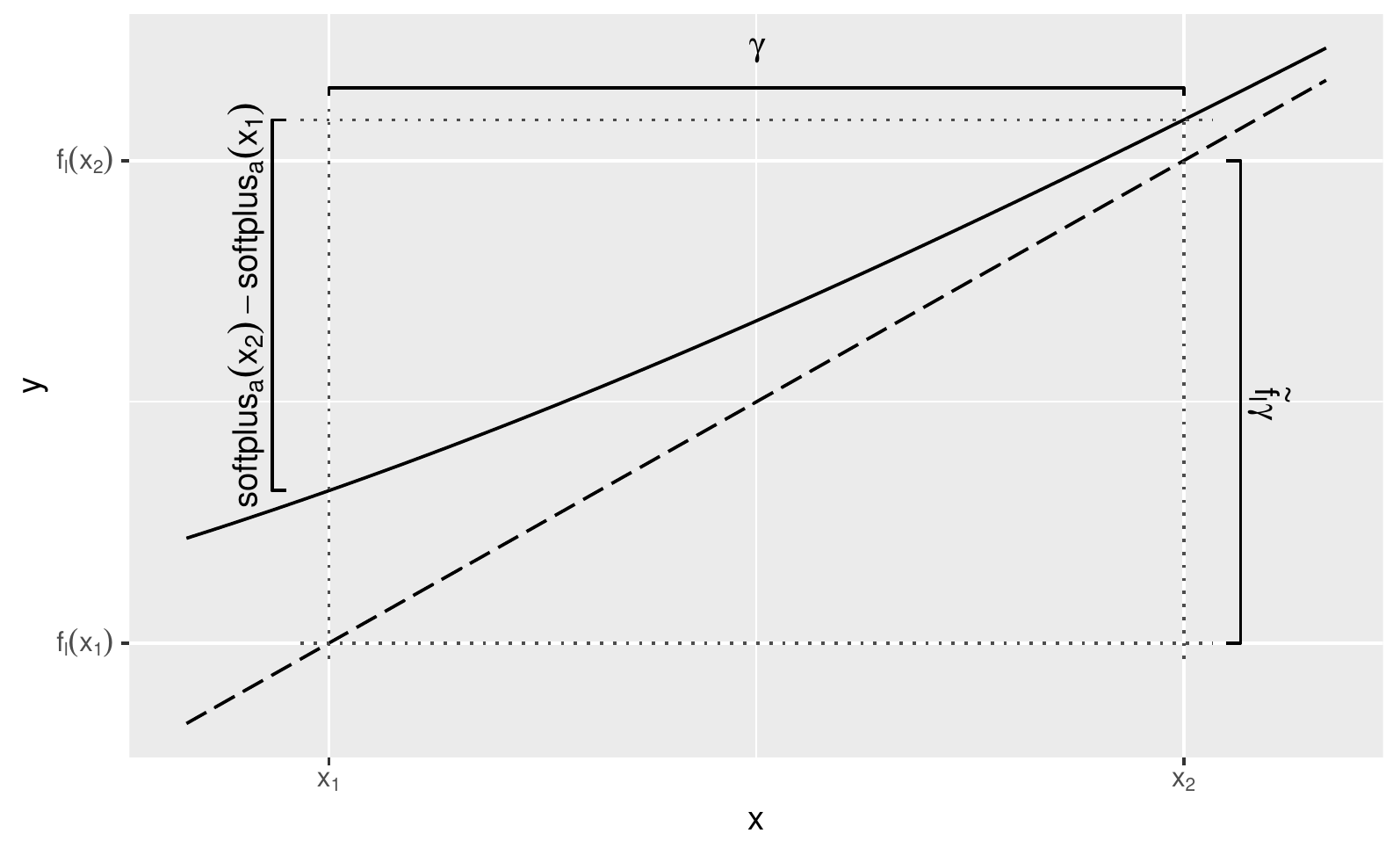}
  \caption{Graphical derivation of the relative error. The solid and dashed line represent the softplus function and the linear function $f_l$, respectively. The error induced by using the softplus function when translating a change of $\gamma$ to the parameter space is given by $f_l(x_2) - f_l(x_1) - (\softplus(x_2) - \softplus(x_1))$. The relative error arrises from taking this quantity relative to $\gamma$.}
  \label{fig:deriv-lin-part}
\end{figure}

Consider a linear function $f_l$ on $\mathbb{R}$ with $\text{d}/\text{d}x\ f_l(x) = \tilde f_l > 0$ for all $x \in \mathbb{R}$.
Then, starting at $x_1 \in \mathbb{R}$ and changing the argument by $\gamma \in \mathbb{R}$ leads to a change under $f_l$ of $\tilde f_l \gamma$, or in other words, $f_l(x_2) - f_l(x_1) = \tilde f_l\gamma$  with $x_2 = x_1 + \gamma$.
See Figure~\ref{fig:deriv-lin-part} for a graphical representation.
This change under $l$ is, of course, constant over the whole domain of $f_l$.
When used as a response function with $\tilde f_l \neq 0$, a regression effect of size~$\gamma$ can be interpreted as influencing the parameter by $\tilde f_l\gamma$ for any $x_1 \in \mathbb{R}$.
If $f_l$ is the identity function, we have $\tilde f_l = 1$ and consequently a change in the predictor equals the same change on the parameter modeled.

More care has to be taken when considering the softplus function since the same change in the argument lead to different changes under the softplus function depending on the location of $x_1$.
We want to find the region for which we can interpret the change in the predictor as if it would change the parameter under $f_l$.
Therefore, we need to assess the error induced by the softplus function and define an acceptable error threshold.
Figure~\ref{fig:deriv-lin-part} displays the different changes under $f_l$ and $\softplus_a$.
The error induced by using the softplus function instead of $f_l$ is given by
\begin{align}
\operatorname{error}_a(x_1, x_2, \tilde f_l)
    = (x_2 - x_1)\tilde f_l - (\softplus_a(x_2) - \softplus_a(x_1)).\label{eq:sp-error}
\end{align}
Considering this error relative to the change under $f_l$, we obtain the relative error
\begin{align*}
  \operatorname{rerr}_a(x_1, x_2, \tilde{f}_l)
    &= \frac{(x_2 - x_1)\tilde f_l - (\softplus_a(x_2) - \softplus_a(x_1))}{(x_2 - x_1)\tilde f_l}\\
    &= 1 - \frac{\softplus_a(x_2) - \softplus_a(x_1)}{(x_2 - x_1)\tilde f_l}.
\end{align*}

Another way to derive the error induced by the softplus function, is to consider the definite integral of the first derivative.
The change under $f_l$, i.e. $f_l(x_2) - f_l(x_1)$, equals the definite integral from $x_1$ to $x_2$ of the first derivative.
Consequently, one can assess the error as the integral of the absolute deviations between the first derivatives of the linear function and the softplus function, i.e.,
\[
  \int_{x_1}^{x_2} \left| \frac{\text{d}}{\text{d}z} f_l(z) - \frac{\text{d}}{\text{d}z}\softplus_a(z) \right| \text{d}z.
\]
As expected, this approach leads to the same measure as defined in Equation~\eqref{eq:sp-error}.
In this paper, we are interested in the interpretation w.r.t. the identity function, thus the relative error is given by
\begin{align*}
  \operatorname{rerr}_a(x_1, x_2)
    &= 1 - \frac{\softplus_a(x_2) - \softplus_a(x_1)}{x_2 - x_1}.
\end{align*}
We say that interpreting a regression effect $\gamma$ directly on the parameter is valid if, for some pre-specified acceptable relative error $\alpha$, the predictor $\eta$ is in the interval $[T, \infty) \subseteq \mathbb{R}$ for which $\operatorname{rerr}_a(T, T + \gamma) < \alpha$ holds.
The acceptable relative error, of course, depends on the application and should be chosen carefully.
In this paper, we consider a relative error of 5\% acceptable.

\subsection{Further properties of the softplus function}

The softplus function shares a number of its properties with the exponential function.
Both functions are smooth and bijective mappings from the real numbers to the positive half-axis.
The first derivative of the softplus function is always positive and is given by
\begin{align*}
  0 < \frac{\mathrm d}{\mathrm dx}\softplus_a(x) = \frac{1}{1 + \exp(-ax)} < 1.
\end{align*}
The second derivative is likewise strictly positive
\begin{align*}
  0 < \frac{\mathrm d^2}{\mathrm dx^2}\softplus_a(x) = \frac{a \exp(ax)}{(1 + \exp(ax))^2}.
\end{align*}
Therefore, the softplus function is strictly monotonically increasing and strictly convex.

\subsection{Inference}\label{sec:inference}

Replacing the standard exponential response function with the softplus response functions introduced in this paper does not cause major difficulties as long as the parameter $a$ is fixed.
Since the softplus-based response functions are continuously differentiable, standard maximum likelihood inference can be used in GLM-type setting where only the derivative of the link function in the definition of the working weights and the working observations of iteratively weighted least squares (IWLS) optimization have to be replaced \citep[see for example][for details on the IWLS algorithm]{fahrmeirRegression2013}.

In our simulations and applications we rely on the Bayesian paradigm for statistical inference, since this allows us to apply the softplus-based response functions also beyond GLMs, for example in generalized structured additive regression models with complex additive predictor \citep{brezgerGeneralizedStructuredAdditive2006} or in structured additive distributional regression models \citep{kleinBayesianGeneralizedAdditive2015}.
For the case of pre-specified response function with parameter $a$, we rely on an MCMC simulation scheme where we update the parameter vector block-wise with a Metropolis-Hastings (MH) step in conjunction with IWLS proposals \citep{gamermanSamplingPosteriorDistribution1997,kleinBayesianGeneralizedAdditive2015}.
IWLS proposals automatically adapt the proposal distribution to the full conditional distribution and therefore avoid manual tuning which is, for example, required in random walk proposals.
This is achieved by approximating the full conditional distribution with a multivariate normal distribution whose expectation and covariance matrix match mode and curvature of the full conditional distribution at the current state of the chain which can be determined based on the IWLS algorithm of frequentist maximum likelihood estimation without requiring the normalizing constant of the full posterior.
More precisely, the parameters of the proposal distribution are determined by executing one Fisher-Scoring step and using the new position as the mean for the multivariate normal distribution while the covariance of the normal distribution is set to be the inverted observed Fisher-Information at the old position.
More formally, let $\thetavec$ be the vector of parameters that should be updated within a MH-block and let $\mathcal{L}(\thetavec)$ be the unnormalized full conditional posterior log density with respect to the parameter vector $\thetavec$.
The proposal distribution is Normal with mean $\muvec = \thetavec + \gvec \Fmat^{-1}$ and covariance matrix $\bm{\Sigma} = \Fmat^{-1}$ where $\gvec$ denotes the gradient of $\mathcal{L}(\thetavec)$ and $\Fmat$ denotes the Hessian of $-\mathcal{L}(\thetavec)$ each with respect to $\thetavec$.
This sampling scheme has proven to be effective in various regression models \citep{langBayesianPSplines2004, kleinBayesianGeneralizedAdditive2015, kleinSimultaneousInferenceStructured2016}.
Our implementation relies on an extension of the R-package \texttt{bamlss} \citep{umlaufBAMLSSBayesianAdditive2018} which implements methodology described above.

\section{Simulations}\label{sec:simulation}

With our simulations, we
\begin{itemize}
 \item conduct a proof of concept evaluation that investigates how reliable models with the softplus response function can be estimated and whether the resulting credible intervals are well calibrated,
 \item study the ability of model selection criteria to distinguish between data generating processes involving either the softplus or the exponential response function, and
\end{itemize}
For all simulations, estimation is conducted within the Bayesian paradigm and carried out in \texttt{R} \citep{rcoreteamLanguageEnvironmentStatistical2019} with the package \texttt{bamlss} \citep{umlaufBAMLSSBayesianAdditive2018}.
We use a similar data generating process varying only the sample size and the response function.
In particular, we assume that the data are generated from a Poisson distribution with expectation $\operatorname{E}(y_i) = \lambda_i = \operatorname{h}(\eta_i)$ where $\operatorname{h}$ denotes the response function.
For a single observation, we choose the predictor structure $\eta = 1.0 + 0.5 x_1 + 1.0 x_2 + 2.0 x_3$ with $x_1, x_2, x_3$ being independent and identically uniform distributed on the interval from $-1$ to $1$. All observations are simulated as being stochastically independent.
Throughout this section, we assume flat priors for all regression coefficients.

\subsection{Point Estimates and Credible Intervals}\label{simsec:cis}

In the first part of the simulation studies, we show that the softplus function can be reliably used as a response function and that posterior means and credible intervals are well-calibrated. The simulation scenarios feature the sample sizes $n \in \{50, 100, 200, 500, 1000,\allowbreak 5000\}$ and the softplus parameter $a$ was set to a value from $\{1, 5, 10\}$. Within each scenario, we simulated 6150~replications.
The number~6150 is determined from considering the coverage of the true parameter as a Bernoulli experiment and requiring that the normal approximation of the 95\% confidence interval for a coverage rate of $0.8$ is smaller than $0.02$.
We run one MCMC chain with 12000 iterations of which the first 2000 iterations are considered as burnin phase.

\begin{figure}[tb]\centering
  \includegraphics[width=\textwidth,keepaspectratio]{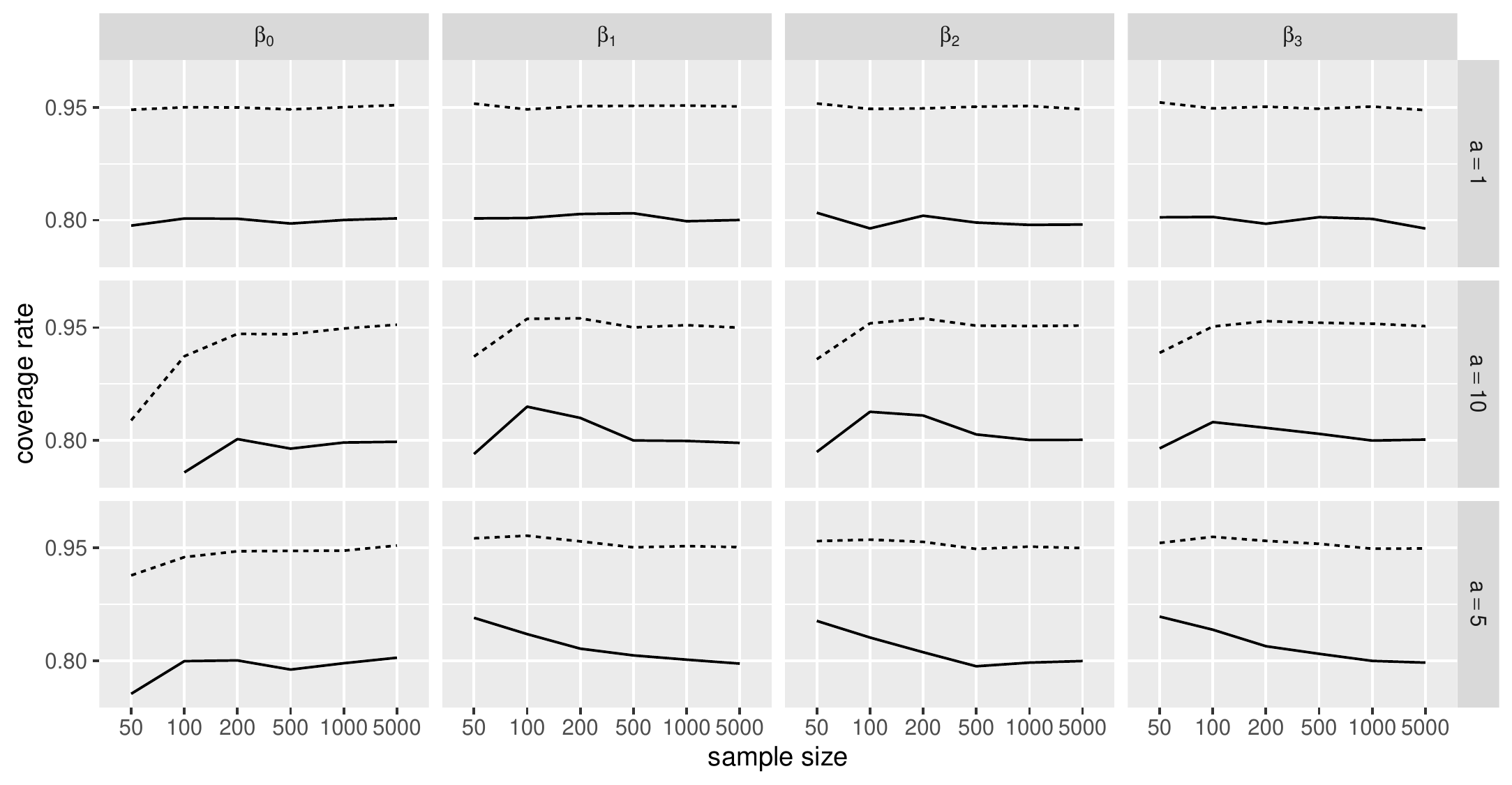}
  \caption{
    \label{fig:cis}
    Coverage probability for 80\% (solid line) 95\% (dotted line) credible intervals for different sample sizes and different softplus parameters $a$.
  }
\end{figure}

To visualize the results, we show box plots of the posterior mean estimates in Figure~\ref{fig:consistency}, and plot coverage rages of 80\% and 95\% credible intervals in Figure~\ref{fig:cis}.
In summary, we draw the following conclusions:
\begin{figure}[tb]\centering
  \includegraphics[width=\textwidth,keepaspectratio]{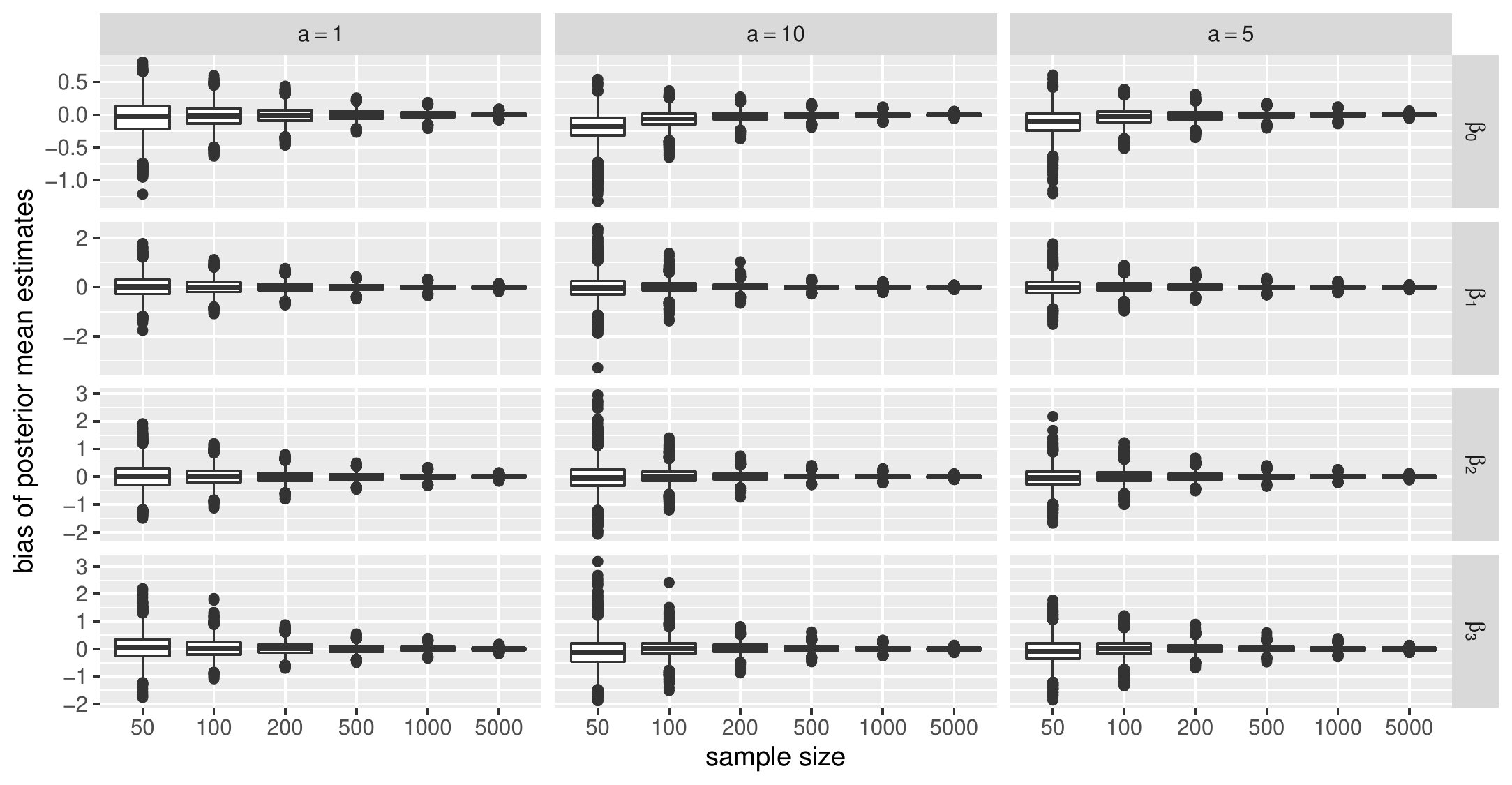}
  \caption{
    \label{fig:consistency}
    Box plots of deviations from posterior mean estimates of regression coefficients to the true value for different sample sizes and different softplus parameters $a$. Replications that include an absolute deviation larger than five for one coefficient have been excluded from plotting for better visualization. This applies to one replication with $a = 5$ and to ten replications with $a = 10$ each with a sample size of $50$.
    }
\end{figure}

\begin{description}
  \item[Bias] For most simulation settings, the bias is negligibly small. The only exception is a small sample size in conjunction with a rather large softplus parameter $a$ where we can observe a small bias, especially for the intercept. However, one has to keep in mind that a large parameter $a$ implies an almost linear link function such that there is considerably less variability (and therefore information) in data sets with softplus response function as compared to the exponential response function with the same linear predictor. Furthermore, the softplus function maps even small negative values to a positive value that is close to zero and thus close to the boundary of the parameter space.  The bias quickly diminishes as the sample size increases.
  \item[Coverage rates] Figure~\ref{fig:cis} supports that our Bayesian approach provides valid credible intervals when the number of observations is large enough. For smaller sample sizes, the coverage rates suffer from the bias introduced by using larger values for the softplus parameter.
\end{description}
In short, the results obtained with the softplus response function are reliable.
Especially for larger sample sizes, no biases are observed and the coverage rates behave as expected.
Results of this simulation exercise obtained with maximum likelihood inference are virtually identical and are omitted for the sake of brevity.

\subsection{Model Selection Based on DIC}\label{sec:model_sel_dic}

In this simulation setting, we study how successfully the well established deviance information criterion \citep[DIC,][]{spiegelhalterBayesianMeasuresModel2002} can be used to discriminate between data generated by either the softplus or the exponential response function.
As in the last subsection, we vary the sample size and use $a=1$~or $a=5$~for the softplus parameter.
Each scenario is replicated 500~times and as before, we run one MCMC chain with a burnin phase of 2000~iterations and 10000~sampling iterations.

\begin{figure}[tb]\centering
  \includegraphics[width=\textwidth,keepaspectratio]{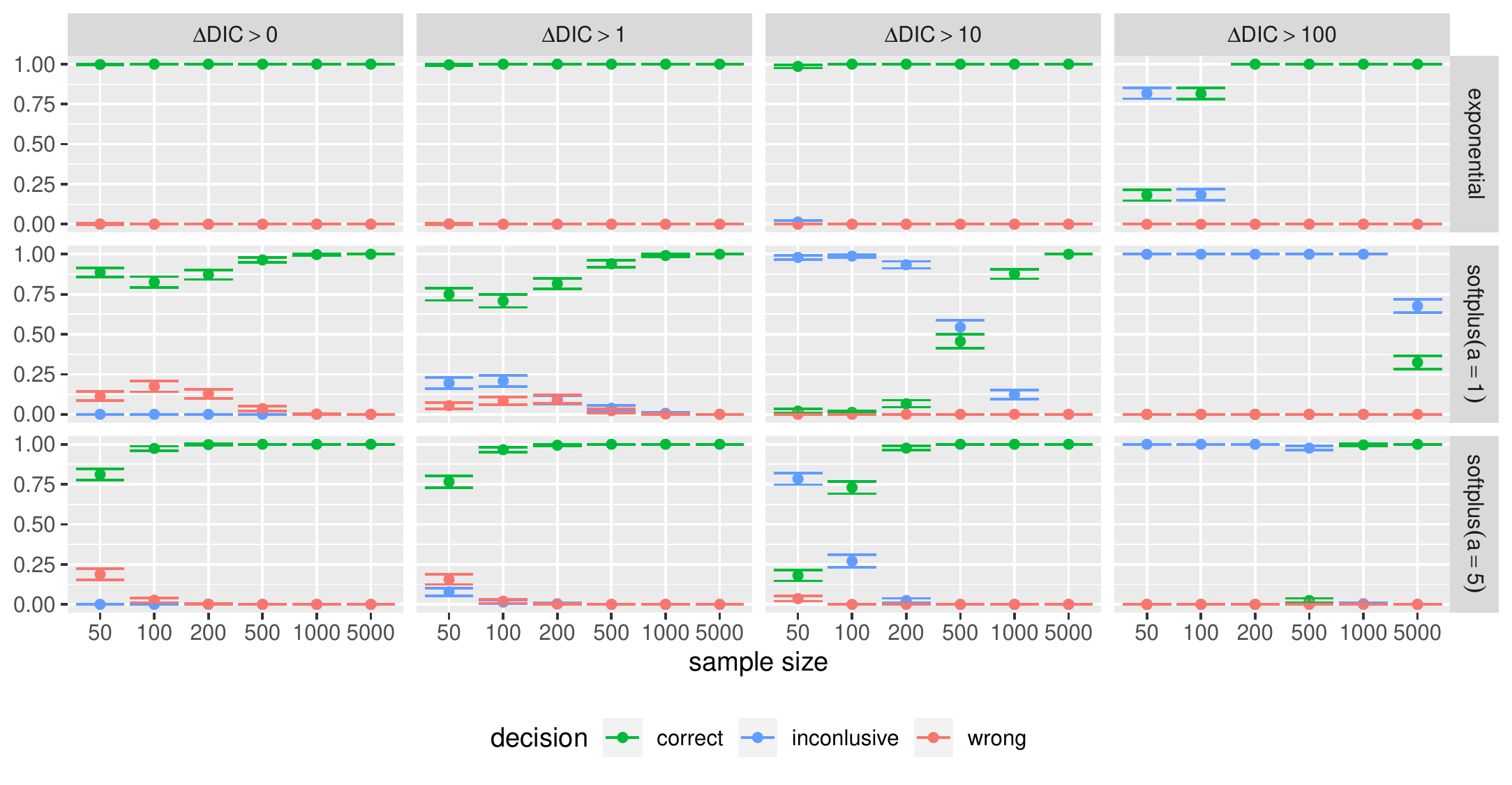}
  \caption{
    \label{fig:selection_dic}
    Percentages of correct model selections based on DIC differences with thresholds 0, 1, 10 and 100 when data are generated with the exponential response function (top row) and the softplus response function (second and third row).
  }
\end{figure}

In Figure~\ref{fig:selection_dic}, we present the results summarized as percentages of correct model selection.
In addition, we consider a more conservative model decision rule where a minimum difference in DIC has to be achieved and use 1, 10, 100 as threshold values.

In all settings, a larger sample size leads to the correct model being recognized more frequently.
Furthermore, the correct model for the same sample size is easier identified if the data are generated with the exponential response function.
As described above, this can be attributed to the fact that the information per observation (quantified by the expected Fisher information) is larger when generated with the exponential response function than with the $\softplus$ response function with $a=5$.
Thus, a larger sample size is needed to have the same probability to select the correct model.
Yet, our simulations show that the DIC is a reliable metric to differentiate between the softplus response function and the exponential response function.

\section{Applications}\label{sec:applications}

We present four applications to demonstrate how the softplus response function can be used in practice.
We contrast our novel approach to the commonly used exponential response function.
First, we employ a well-known data set from ethology about horseshoe crab mating behavior as an illustrative example for count data regression with the softplus response function (Section~\ref{sec:horseshoe_crabs}).
Then, we illustrate the usefullness of the softplus response function in a distributional regression model with smooth effects (Section~\ref{sec:capital_bikes}).
For that, we fit a model to data from a bike-sharing service in Washington D.C., where the softplus function can be used as a response function for the variance parameter of a normally distributed outcome.
Following, we provide details on the motivating example shown in the introduction. Using data from the Munich rent index, we demonstrate similarities between results obtained from quantile regression and the softplus response function (Section~\ref{sec:munich-rents}). 
In an application to operational loss data (Section~\ref{sec:jh-application}), we demonstrate the usefulness of the softplus response function apart from the quasi-additive interpretation.

In the supplementary material, we revisit the horseshoe crab data estimating the limiting gradient of the softplus exponential function which suggests to use a linear response function.
Furthermore, we analyze data from the Australian Health Survey 1977--1987 focusing on the number of physician consultations within two weeks in relation to explanatory variables as another illustrative example.

\subsection{Horseshoe Crabs}\label{sec:horseshoe_crabs}

\cite{brockmannSatelliteMaleGroups1996} investigates horseshoe crab mating behavior.
Horseshoe crabs have a strongly male-biased sex ratio which becomes particularly apparent in spring when male and female horseshoe crabs arrive in pairs at the beach ready to spawn.
Unattached males also come to the beach, gather around females and try to gain fertilization at the expense of the attached males.
\cite{brockmannSatelliteMaleGroups1996} shows that the number of unattached males, so-called satellites, gathering around a couple depends mainly on properties of the female crab, and to a lesser extent on environmental factors.

\cite{agrestiCategoricalDataAnalysis2013} and \cite{kleiberVisualizingCountData2016} reanalyze these data using count data regression techniques to model the number of satellite males.
\cite{agrestiCategoricalDataAnalysis2013} assumes the response to be Poisson or negative binomial distributed and for each response distribution he compares the exponential response function and the identity response function.
Among these four models, he finds that the negative binomial regression model with identity response function fits the data best.
\cite{kleiberVisualizingCountData2016} extend this approach by using hurdle models to allow excess zeros.
The authors favor the negative binomial hurdle model with exponential response function.
However, they omit results for the identity response function since they claim that the negative binomial hurdle model is superior with respect to predictions compared to the negative binomial model with identity response function favored by \cite{agrestiCategoricalDataAnalysis2013}.
Their argumentation includes that, in contrast to the identity response function, the exponential response function avoids negative predictions for small carapace widths.
The softplus function prevents negative predictions as well.

To illustrate how the softplus function can be used to model the bounded expectation of a count data model, we extend the analyses mentioned.
For that, we use the softplus function with $a=5$ as a response function in negative binomial regression models with and without accounting for excess zeros.
Following \cite{kleiberVisualizingCountData2016}, the carapace width and a numeric coding of the color variable are used as regressors in all models.
All models are fitted with \texttt{bamlss} using uninformative priors on all coefficients.

To compare the relative performances of the eight models, we use the DIC (see Table~\ref{tab:hs-dic}).
Similar to \cite{kleiberVisualizingCountData2016}, we find that the negative binomial hurdle models fit best and the DIC slightly favours the softplus response function.
The small difference in fit between the response functions is not surprising since \cite{kleiberVisualizingCountData2016} already point out that, given at least one satellite, neither carapace width nor color seem to have a significant contribution.
Note, an intercept-only model does not depend on the response function used since the intercept parameter can adapt to the response function yielding the same distribution parameter.
Consequently, the limited impact of the response function in the zero-adjusted model is expected.
\input{arxiv-files/hs_dic_tab}

\begin{figure}[hbt]
  \centering
  \includegraphics[width=1\textwidth]{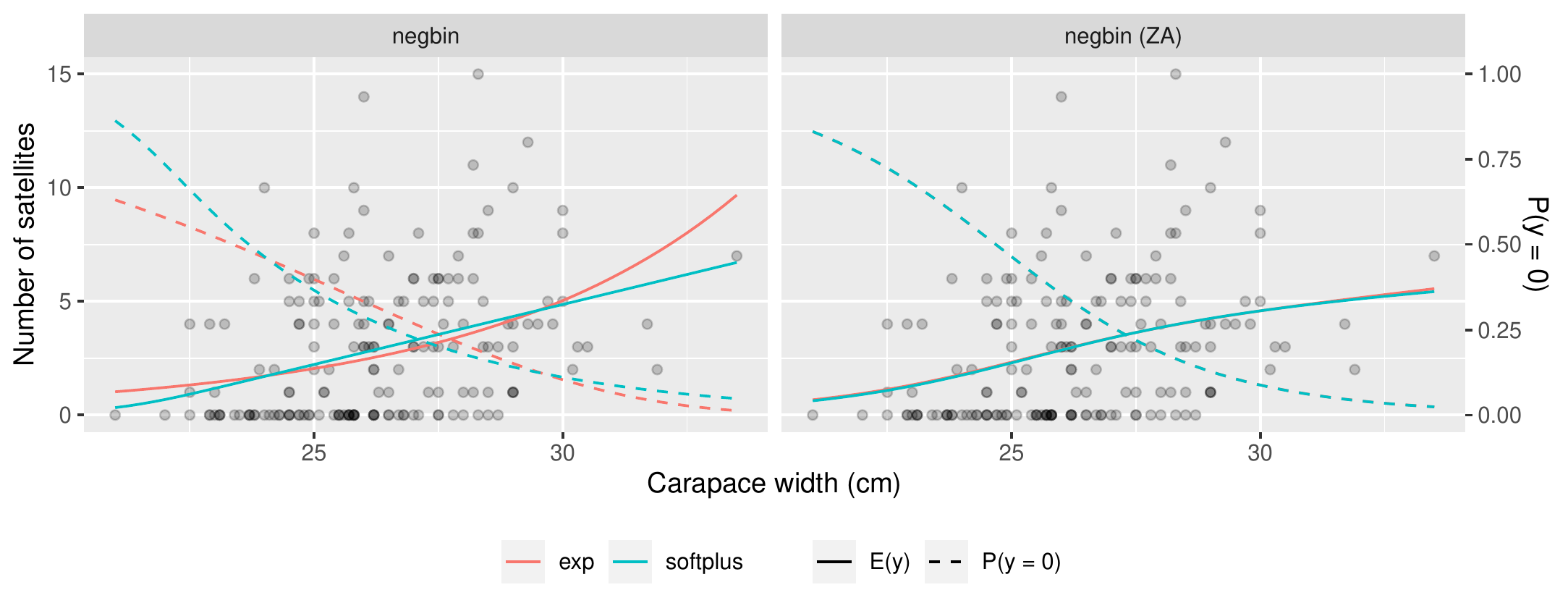}
  \caption{Plots of the expected response given carapace width and mean color for different response distributions broken down by response function. ZA indicates the the zero-adjusted response distribution. In addition, the dashed line indicates the probability of 0 satellites. The points show the observed data. }\label{fig:hs-expectations}
\end{figure}

Nonetheless, the application gives insight to the usefulness of the softplus function.
In Figure~\ref{fig:hs-expectations}, we display the expected number of satellites predicted as a function of carapace width with color set to the mean value.
When considering the negative binomial regression, one can clearly observe the differently shaped curves reflecting the response function employed.
A visual examination suggests that the exponential response function might not decay fast enough for small values of carapace width while increases too fast for large values.
On the contrary, the softplus response function seems to fit better when compared to the pattern arising from the model with zero-adjusted negative binomial response distribution (i.e., the hurdle model).

\begin{figure}[bt]
  \centering
  \includegraphics[width=\textwidth]{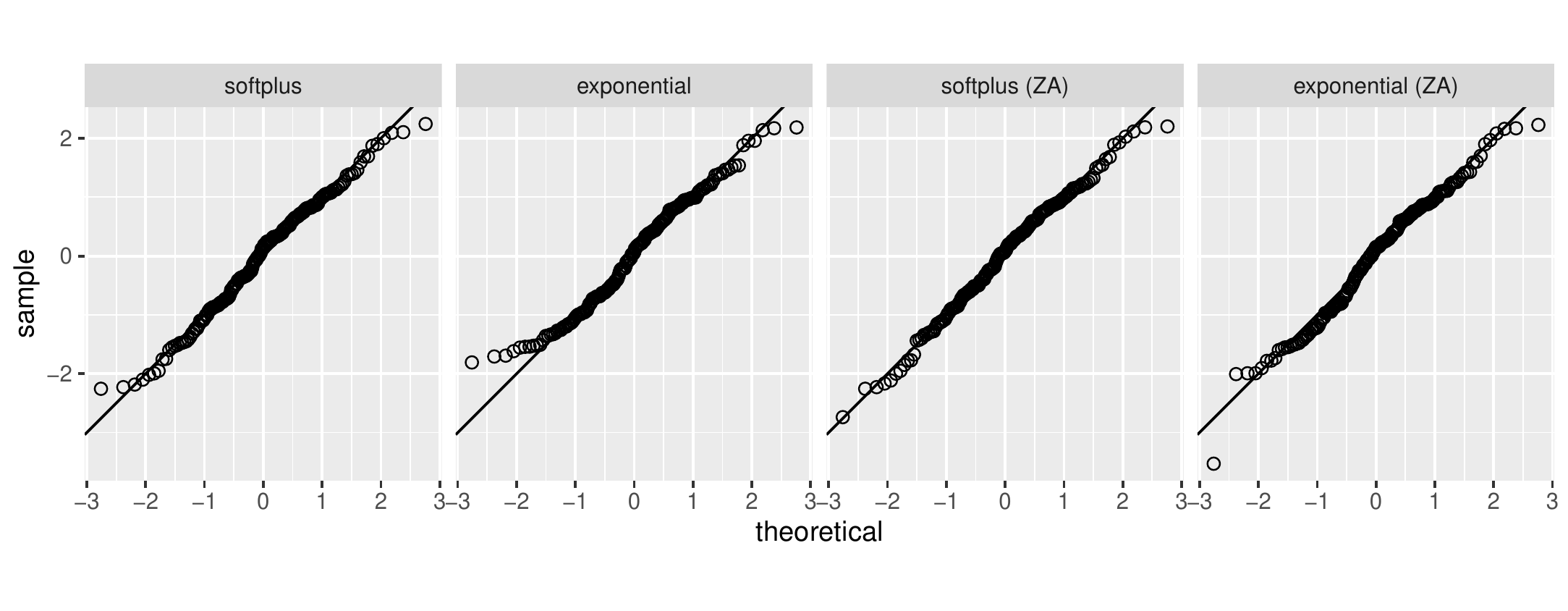}
  \caption{QQ-plots of one realization of RQRs for negative binomial distributed responses without and with zero-adjustment (indicated as ZA) employing the softplus or the exponential response function.}
  \label{fig:hs-qq}
\end{figure}

In particular, when considering the probabilities of observing zero satellites ($\operatorname{P}(y = 0)$; these are represented as dashed lines in Figure~\ref{fig:hs-expectations}), the model based on the softplus function is closer the output from the zero-adjusted response distribution.
This is especially true for small width values of the carapace.
This is due to the fact that the softplus function with $a = 5$ approaches zero much faster than the exponential response function does.
Furthermore, quantile-quantile plots (QQ-plots) of the randomized quantile residuals \citep[RQRs,][]{dunnRandomizedQuantileResiduals1996} indicate a decent fit to the data for all models with a preference for the hurdle model (see Figure~\ref{fig:hs-qq} for one realization).

When removing the zero-adjusted model from our considerations, the DIC suggests that the model with softplus response function has an advantage.
This finding is in line with \cite{agrestiCategoricalDataAnalysis2013} and his claim of a better fit using the identity response function compared to the exponential response function.
By fitting Poisson models with softplus response function and exponential response function, we can confirm the results from \citet{agrestiCategoricalDataAnalysis2013}, i.e., the quasi-linear response function fits the data better in terms of DIC.
However, we omit the results here because \citet{kleiberVisualizingCountData2016} have already pointed out that the Poisson response distribution can not appropriately model the data.

\input{arxiv-files/hs_coef_tab}

To illustrate the difference in the interpretation of softplus and exponential response function, we focus on the model assuming a negative binomial distributed response without adjusting for zeros since the impact of the different link functions becomes almost indistinguishable when adjusting for zeros.
Posterior means of the parameters are displayed in Table~\ref{tab:hs-coef} together with the corresponding 95\% credible interval (equal-tailed).
For a change of $0.53$, the linear threshold, as defined in Section~\ref{sec:linear-part}, is $0.37$, while for a change of $-0.54$, its value is $0.91$.
Notice that more than $98\%$ and $94\%$ of the posterior means of the linear predictor are larger than these linear thresholds.
Thus, we consider the linear interpretation of the covariate effects of width and color as valid for almost all observation.
In particular, a change by one unit in carapace size or color would increase the expected number of satellites by $0.53$ or $-0.54$, respectively.
This is in contrast to the interpretation of the exponential response function where the same changes would lead to a multiplicative change of $1.20$ and $0.77$, respectively.
The 95\% credible interval of the effect of color includes $0$ for the softplus response function but not for the exponential response function.
In both cases, however, the null effect is very close to the credible interval's boundary.

\subsection{CapitalBikeshare}\label{sec:capital_bikes}
In this section, we demonstrate the applicability of the softplus function as a response function in a Bayesian distributional regression model with flexible covariate effects.
We employ data from CapitalBikeshare, a bicycle sharing service located in Washington D.C., to analyze the mean rental duration in minutes within each hour in the years 2016 -- 2017\footnote{the raw data can be found at https://www.capitalbikeshare.com/system-data}.
The operator might want predict the number of trips and their expected duration in order to know how many bikes have to be stocked.
However, the variance of the average journey time is also important, as it can prevent bottlenecks caused by unforeseen fluctuations.

The data have been preprocessed by the following rules:
\begin{itemize}
  \item Trips taken by staff for service and inspection of the system have been removed as well as trips towards test stations.
  \item Trips taken by non-members have been removed.
  \item All trips with a duration of fewer than 60~seconds have been removed since they most likely indicate a false start or users ensuring that the bike is secure by redocking it.
  \item Trips longer or equal to 60 minutes have been removed. This amounts to roughly 0.5\%~of the eligible trips. We consider them as outliers since the financial incentive system of CapitalBikeshare strongly encourages users to return bikes within the first hour.
\end{itemize}

The mean rental duration per hour is on average based on $308.24$~trips.
A raw descriptive analysis of this quantity gives an average of $10.9$~minutes with a standard deviation of $1.88$~minutes.

The framework of structured additive distributional regression models \citep{rigbyGeneralizedAdditiveModels2005,umlaufBAMLSSBayesianAdditive2018} extends the generalized additive models such that multiple parameters of a response distribution can be modeled with structured additive predictors and a suitable response functions.
For our analysis, we assume the mean rental duration to be conditionally independent and normally distributed.
We model both distributional parameters (mean and standard deviation) with structured additive predictors.
In particular, the mean rental duration within each hour $y_i$ is assumed to be independently and normally distributed with mean $\mu_i$ and standard deviation $\sigma_i$.
The parameters are linked to predictors ($\eta^{\mu}_i$, $\eta^{\sigma}_i$) via response functions $\h^{\mu}$ and $\h^{\sigma}$.

We use the same structure for both predictors and drop in the following superscript index.
The predictor is specified as $$\eta_i = f_1(\texttt{yday}_i) + f_2(\texttt{dhour}_i) + \xvec_i'\betavec,$$ where \texttt{yday} denotes the day of the year, \texttt{dhour} denotes the hour of the day and the  last term contains the intercept and additional linear effects.
As linear effects, we consider a dummy variable for the year 2017 and a binary variable that encodes if the trip took place on a weekend.
The smooth functions are represented by cyclic P-splines \citep{eilersFlexibleSmoothingBsplines1996,hofnerUnifiedFrameworkConstrained2016} with second-order random walk penalty \citep{langBayesianPSplines2004}.

To illustrate the difference in interpretation between the softplus response function and the popular exponential response function, we estimate the model for both response functions, i.e.\ $\h^{\sigma} = \exp$ or $\h^{\sigma} = \softplus_{10}$.
The DIC favors the softplus response function (exponential: $58152$, softplus: $57943$).
The softplus parameter was not chosen on the basis of an information criterion, but rather to enable the quasi-additive interpretation.

Detailed results concerning the mean predictor and its components are omitted since both models employ the same response functions and the results are very similar (see the Supplementary Material for a full description).

\begin{figure}[t]\centering
	\includegraphics[width=\textwidth]{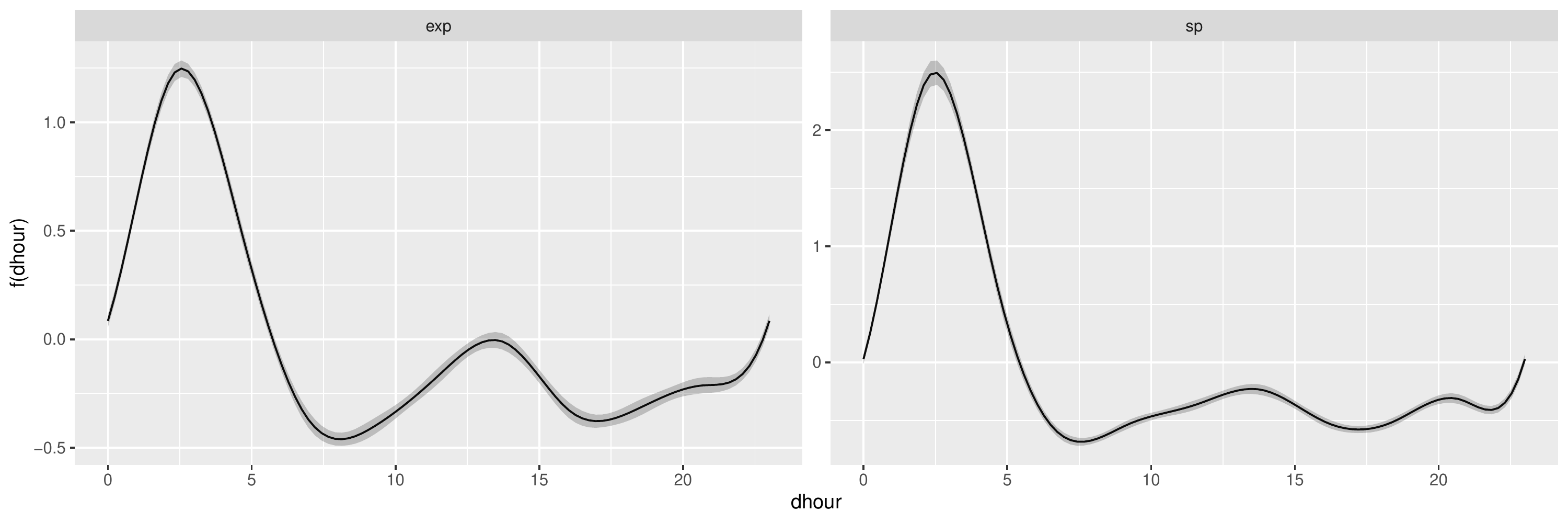}
	\caption{\label{fig:cbs-sigma-dhour} Posterior mean estimates on the predictor of the standard deviation together with 95\% point-wise credible intervals (equal-tailed) for both response functions.
	}
\end{figure}

We focus on the effect of the response function with respect to the standard deviation $\sigma_i$
and, in particular, on the smooth effect of \texttt{dhour} and the linear effect of \texttt{weekend}.
Figure~\ref{fig:cbs-sigma-dhour} shows the estimated effect of the time of the day on the predictor of the standard deviation.
We find that both models yield similar patterns.
The standard deviation is much larger in the early hours of the day with a peak around 3~am, then drops steeply, crosses the zero line shortly after 5~am and is comparatively low in the morning.
Over the course of the morning the standard deviation increases slightly until lunchtime, then decreases over the early afternoon before starting to increase again around 4~pm. At first slightly and then very steep until it reaches its peak again in the early morning hours.

The direct interpretation of these effects is difficult, especially when using the exponential response function.
For the softplus model, the estimated values of the linear predictor are larger than $0.42$ and in conjunction with a softplus parameter of 10, covariate effects can be interpreted as quasi-additive effects on the parameter (the relative error for a change of $0.0001$ at predictor value $0.42$ is smaller than $2\%$).
In the following, consider the difference between the initial peak at 2.5~am and the second peak at lunch time.
In the model with the softplus response function, we observe that the predictor decreases by about 2.7~units and, consequently, the standard deviation likewise.
In contrast, for the competing model, the exponential function must be applied to the predictor, and the outcome can subsequently be interpreted multiplicatively.
The exponential model outputs an additive change of $-1.25$ on the predictor $\eta^{\sigma}$, which is reflected in a multiplicative change of the standard deviation by $3.5^{-1}$.

When considering the variable \texttt{weekend} (Table~\ref{tab:cbs-sigma-le}), its effect is similar in both models: the mean rental duration exhibits more variance during the weekend, and both 95\% credible intervals exclude zero.
However, the interpretation of the exponential model is not straightforward: the posterior mean of the  regression coefficient related to \texttt{weekend} is 0.39.
In order to assess the multiplicative effect of weekend on the standard deviation, one needs to consider the posterior mean of the transformed parameter, that is  $\overline{\exp(\beta_\text{weekend})} = 1.48$.
We conclude that on a weekend the standard deviation is 1.48~times larger than on weekdays.

The softplus model directly outputs the additive effect of weekend.
We expect that the standard deviation of the mean rental duration is $0.56$~minutes larger on weekends than on working days.
\input{arxiv-files/cbs-sigma-le}

\begin{figure}[tb]
  \centering
  \includegraphics[width=\textwidth]{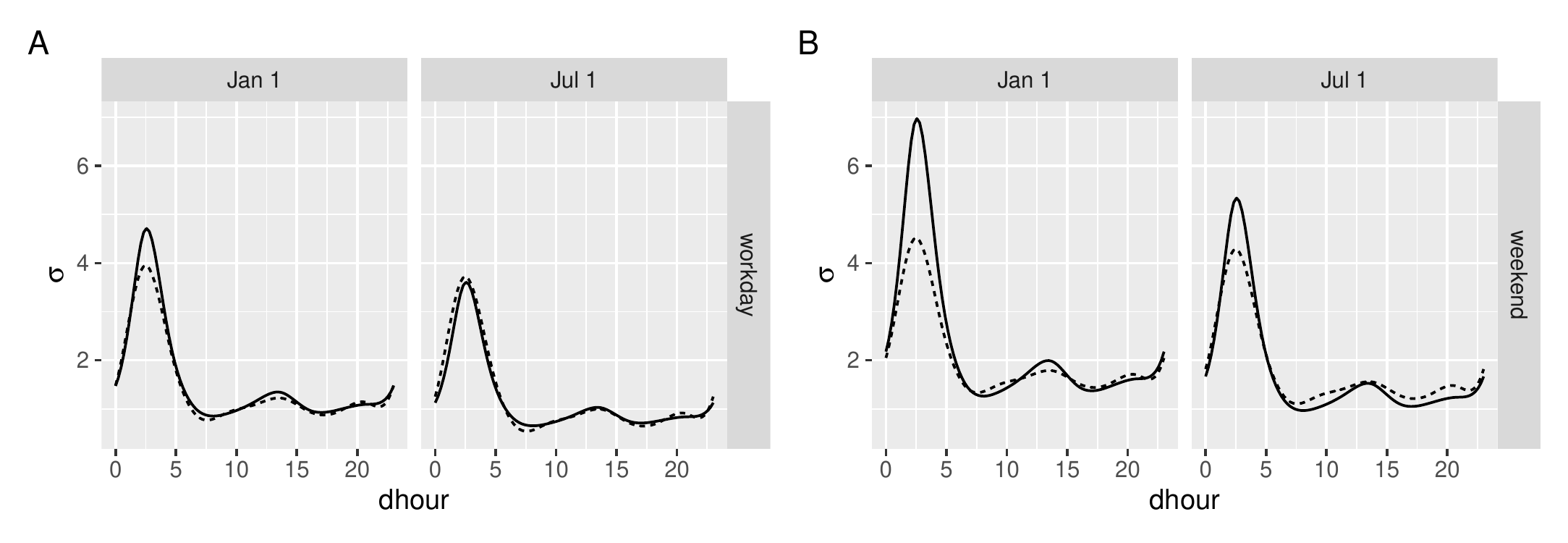}
  \caption{Predicted standard deviation over the course of a day. Exemplary for the first day of the year~2016 and July 1st on a working day (Panel~A) and on a weekend (Panel~B). The solid line refers to predictions for the model with exponential response function while the dashed line refers to the model with softplus response function.}
  \label{fig:cbs-pred-sigma}
\end{figure}

In both models, the interpretation of regression effects w.r.t.\  the predictor $\eta^{\sigma}$ is straightforward and the nature of the effects w.r.t.\ the standard deviation is known (i.e., additive or multiplicative).
Despite this, the combination of effects and their interaction with the response function makes an assessment of the absolute effects on the distribution parameter difficult.
However, it becomes more apparent when we consider plots of the predicted parameter values.
In Figure~\ref{fig:cbs-pred-sigma}, we show the predicted values (using the posterior mean of the estimated parameters) for $\sigma$ over the course of two selected days of the year 2016 (these are 1st of January and 1st of July).
We further add the effect of \texttt{weekend} and display the predicted values for both models.
We observe that both models output similar values for a weekday on the first of July. Even the spike in the early morning appears similar.
In the winter or on a weekend, the standard deviation is larger in both models (exponential model: 1.48~times larger on weekends, 1.31~times larger on the 1st of January; softplus model: 0.56~minutes larger on weekends, 0.23~minutes larger on the 1st of January).
The difference between the models is most apparent at the 3am peak where it is now about one minute.
The almost explosive behaviour of the exponential function becomes apparent when now considering the combined effect (left panel in Figure~\ref{fig:cbs-sigma-dhour}~B).
The difference between the peak at noon and in the morning is almost 5 minutes with the exponential function (that is a 3.5~fold increase) and just 2.75 minutes with the softplus function.
Again, for the remaining time of the day both models output relatively small differences.
Compared to the right panel in~A, we find $\exp(0.27 + 0.39 + 1.25) = 6.75$~fold increase between noon and morning peak with the exponential function. Due to the additive nature of the softplus function, the effects are not multiplied and the difference is just $4.53$~fold (4.49~minutes compared to 0.99~minutes).

From this application, we conclude that the softplus function can be used as an alternative response function in distributional regression for a bounded parameter.
Using the softplus function, can lead to a better model fit.
In addition, due to the limited growth rate of the softplus function it can the avoid explosive behaviour of the exponential function.
Besides, it can offer the quasi-additive interpretation of regression effects.
The softplus function is even a viable alternative if both functions fit equally well and it is up to the practitioner to decide which one to prefer.

\subsection{Munich Rent Index}\label{sec:munich-rents}

In Munich, tenants have to pay some of the highest prices per square meter for housing across Germany.
To provide tenants and landlords with reference prices, the city has established a rent index. %
Using data with 3082~observations from the Munich rent index available on the website accompanying \citet{fahrmeirRegression2013},
we relate the rent in Euros to the size in square meters of the living area.

The data shows significant heteroskedasticity, indicated by increasing variance in the response as the size of the flat increases (see Figure~\ref{fig:munich-rent}).
Therefore, a simple linear model with constant variance is inappropriate.
Besides linear quantile regression \citep{koenkerQuantileRegression2005}, we additionally use two versions of location-scale models within the framework of distributional regression relating the standard derivation to the covariate.
Since the standard derivation is strictly positive, a response function enforcing positivity of the transformed predictor should be used.

We compare the performance of the location-scale models equipped with the softplus response function (with $a$ fixed to 30) and exponential response function assuming normally distributed responses.
The linear predictors for the expectation $\beta_0 + \beta_1 \texttt{area}_i$ and the standard deviation $\gamma_0 + \gamma_1 \texttt{area}_i$ feature beside the linear effect of area only the intercept.
We use flat priors on the parameters and fit the model with \texttt{bamlss}.
In terms of DIC, we find the model with softplus response function fits the data better than the competing model (39281 vs.\ 39301).
A possible explanation seems to be the overestimated variance of the response for flats larger than approximately 100~square meters.
We draw this conclusion by visually comparing the estimated quantiles of the response to the data shown in Figure~\ref{fig:munich-rent}.

Additionally, the figure highlights the explosive behavior of the exponential function for larger arguments which the softplus function avoids.
In particular, the softplus function translates the linear effects in the predictor to the estimated quantiles since the quantile function for a normal random variable with mean $\mu$ and standard derivation $\sigma$ is linear in the standard derivation
\[
  F^{-1}(p) = \mu + \sqrt(2) \sigma \operatorname{erf}^{-1}(2p - 1)
\] where $\text{erf}^{-1}$ denotes the inverse error function.

We further observe quantiles estimated using the softplus response function are roughly comparable to the results obtained via quantile regression \citep{koenkerQuantregQuantileRegression2021}.
However, the estimated quantiles depend on the assumed response distribution.
Here, this is the normal distribution and, therefore the quantiles are symmetric w.r.t.\ the median.
The non-parametric quantile regression model avoids this restriction since it requires no distribution assumption.
However, using a parametric regression model has the advantage over non-parametric quantile regression of preventing undesirable artifacts such as quantile crossing (altough we do not observe this in our application due to the relatively simple model structure that we employ).

\subsection{Operational Losses at UniCredit}\label{sec:jh-application}

In this section, we demonstrate the usefulness of the softplus function in the context of a distributional regression model.
To do so, we employ the data used in \cite{hambuckers2018} where the authors model the size distribution of extreme operational losses in a large Italian bank (i.e., losses stemming from fraud, computer crashes or human errors) given a set of economic factors (e.g., interest rates, market volatility or unemployment rate).
This conditional distribution is then used to estimate a quantile at the 99.9\% level, a quantity needed to establish the regulatory capital held by the bank, with large quantile values requesting more capital, and to monitor operational risk exposure in various economic situations, such as a a financial crisis or economic expansion periods.

Since operational loss data are heavy-tailed and that the focus is on extreme values dynamics, distributional regression techniques are needed to properly reflect the effect of the covariates on extreme quantiles.
Following \cite{chavez2016}, an approach based on extreme value theory is traditionally used: a high threshold~$\tau$ is defined by the statistician, and only losses larger than this threshold are kept for the analysis.
Then, we assume that the distribution of the exceedances above the threshold is well approximated by a Generalized Pareto distribution (GPD).
In the context of extreme value regression, the parameters of the GPD are additionally modeled as functions of covariates, defining a Generalized Pareto (GP) regression model.
Estimated parameters of this model are used to derive the quantile of interest given values of the covariates.

For mathematical and conceptual reasons, both parameters of the GPD are restricted to strictly positive values: the scale parameter $\sigma(x)$ is strictly larger than 0, whereas the shape parameter $\gamma(x)$ is restricted to positive values to guarantee the consistency of the maximum likelihood estimator and to reflect the tail-heaviness of the loss distribution.
Thus, an exponential response function is commonly used for computational simplicity, although no theoretical support for a multiplicative model exists (see, e.g., \cite{umlaufPrimerBayesianDistributional2018, hambuckersQF, bee2019} and \cite{groll2019}).
However, this choice for $\gamma(x)$ might quickly generate explosive quantile estimates for some combinations of the covariates, making the model economically unexploitable to derive capital requirements.
In addition, it can have a similar undesired effect on uncertainty quantification: the width of the confidence interval on the quantile increases exponentially with the estimated quantile itself.
Consequently, it is in times of high estimated risk exposure (i.e., large values of the 99.9\% quantiles) that risk managers face the highest model uncertainty to take decisions.

\begin{figure}[htbp]
  \includegraphics[width=\textwidth]{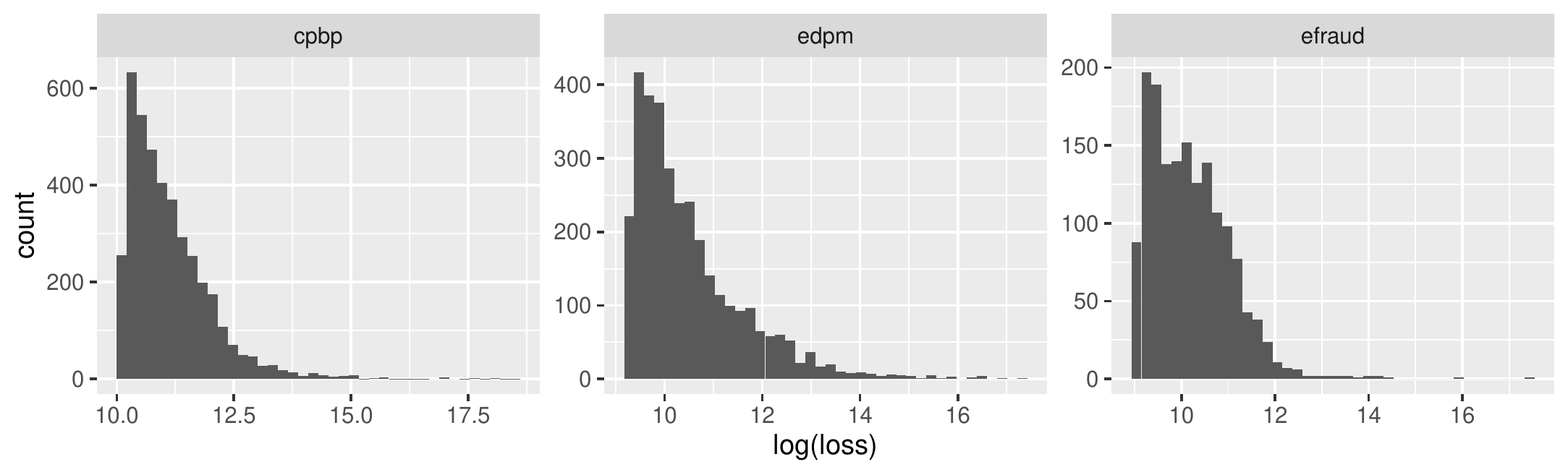}
  \caption{Histograms of the log-losses larger than the 75\% quantile per category, for the three event types.}\label{fig:gpd_histo}
\end{figure}

To illustrate how the softplus function helps mitigating these issues, we reanalyze the UniCredit loss data for three categories of operational losses, namely the categories \textit{execution, delivery and process management} (\texttt{EDPM}), \textit{clients, products, and business practices} (\texttt{CPBP}) and \textit{external fraud} (\texttt{EFRAUD}).
The data were collected over the period January 2004 - June 2014.
As in \cite{hambuckers2018}, we work with the 25\% largest losses in each category. Descriptive statistics and histograms of the data are provided in Table~\ref{tab:ol-desc} and Figure~\ref{fig:gpd_histo}.
They highlight both the presence of extreme values that need to be accounted for. For each loss registered during a given month, we associate the values taken by a set of economic covariates observed the month before and found susceptible to influence the loss distribution by \cite{hambuckers2018} (the complete list can be found in the Supplementary Material).
Denoting by $y_{i}=z_{i}-\tau$ the exceedance of a loss $z_{i}$ above the threshold $\tau$, and by $\mathbf{x}_{\gamma,i}$ and $\mathbf{x}_{\sigma,i}$ the corresponding vectors of covariates for both $\gamma$ and $\sigma$, our model can be written in a generic form as
\begin{align*}
y_{i} &\sim G(\gamma(\mathbf{x}_{i}),\sigma(\mathbf{x}_{i})),\\
\gamma(\mathbf{x}_{i})&=h^{\gamma}(\mathbf{x}_{\gamma,i}^{'}\pmb{\beta}_{\gamma}),\\
\sigma(\mathbf{x}_{i})&=h^{\sigma}(\mathbf{x}_{\sigma,i}^{'}\pmb{\beta}_{\sigma}),
\end{align*}
with $G(\cdot)$ denoting the cumulative distribution function of the GPD, and $\pmb{\beta}_{\gamma}$ and $\pmb{\beta}_{\sigma}$ being the vectors of regression parameters for $\gamma$ and $\sigma$, respectively.

\begin{figure}[tb]
  \includegraphics[width=\textwidth]{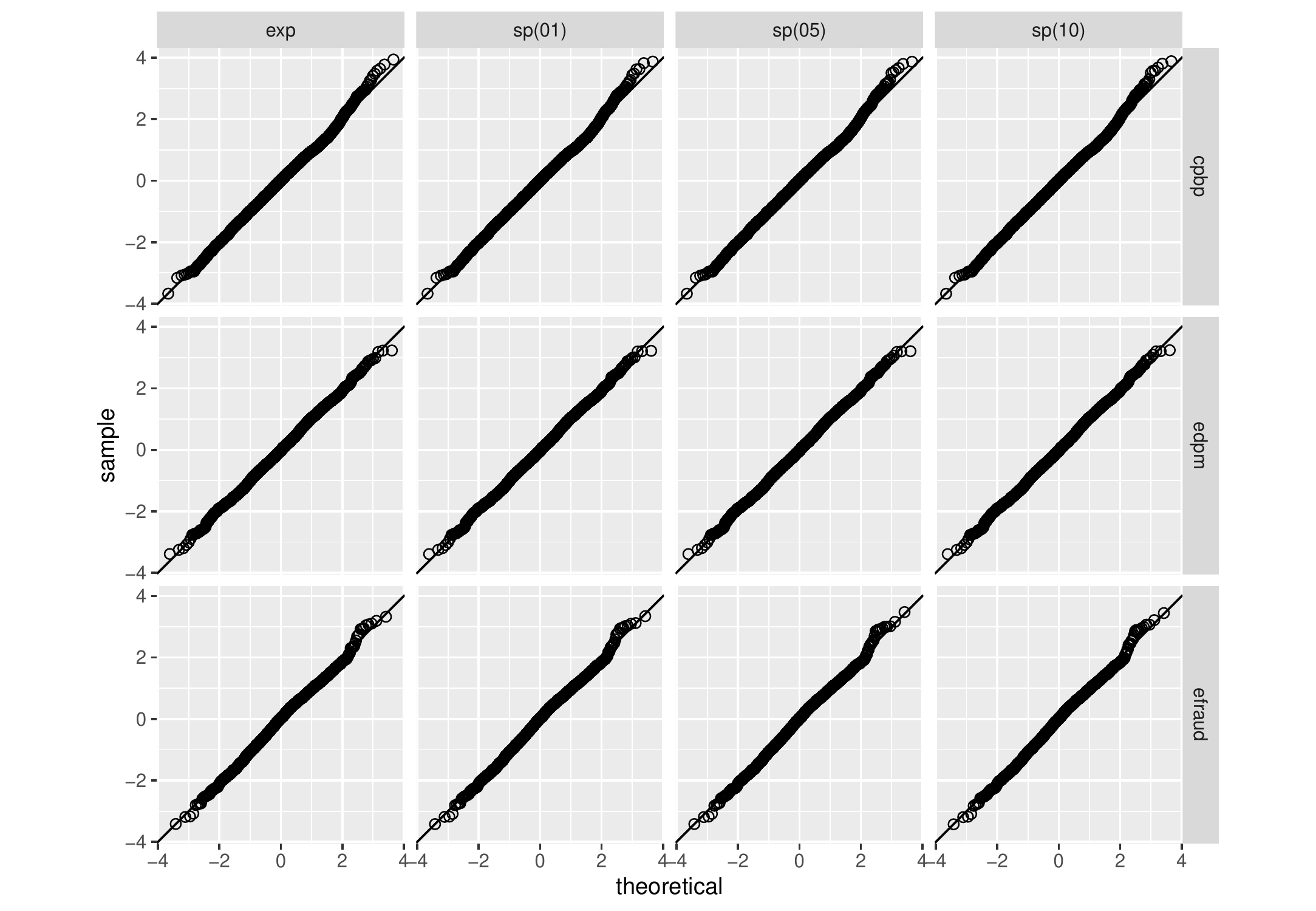}
  \caption{QQ-plots of the RQRs for the different models. The number in parentheses refers to the employed goodness of approximation parameter in the softplus function.}\label{fig:gpd_qqplot}
\end{figure}

We fit separate GP regression models to each sample with \texttt{bamlss} using 24000~MCMC iterations, treating the first 4000~iterations as burn-in and applying a thinning factor of~20.
We compare the results obtained with various response function $h^{\gamma}$ (we keep the exponential function for $h^{\sigma}$). Estimated regression parameters can be found in the Supplementary Material. We report the DIC in Table~\ref{tab:gpd_infocrit}, whereas Figure~\ref{fig:gpd_qqplot} displays the QQ-plots of the RQRs.
They both indicate that the overall goodness-of-fit is satisfactory and similar across models, with a slight preference for the sofplus models for \texttt{EFRAUD}, and an advantage of the exponential model for the other categories.
\input{arxiv-files/ol_desc_stats}

\begin{table}[htbp]
\centering
\begin{small}
\begin{tabular}{l|cccccc}
\toprule
Category & Exp. & sofptplus $1$ & softplus $5$ & softplus $10$ & Null model\\
\midrule
\texttt{CPBP}
 & \underline{23,593.24} & 23,594.99 & 23,598.48 & 23,603.37 & 23,626.18 \\
 \texttt{EDPM}
 & \underline{16,532.26} & 16,532.77 & 16,535.42 & 16,540.51 & 16,542.68\\
 \texttt{EFRAUD}
 & 6,547.11 & \underline{6,545.59} & 6,545.89 & 6,546.31 & 6,567.46\\
 \bottomrule
\end{tabular}
\end{small}
\caption{Deviance information criterion for the different models. \textit{Null model} refers to the exponential model with no covariates. %
}\label{tab:gpd_infocrit}
\end{table}

\begin{figure}[htbp]
\centering
\includegraphics[width=\textwidth]{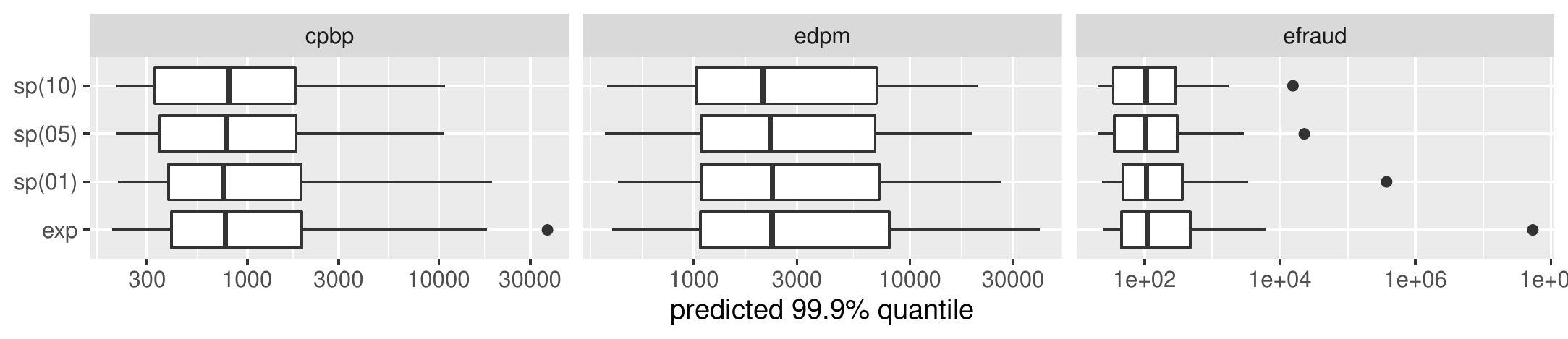}
\caption{Box plots of the estimated 99.9\% quantiles of the conditional loss size distribution (x-axis is in log-scale).}\label{fig:gpd_boxplot}
\end{figure}

\begin{figure}[htbp]
  \includegraphics[width=\textwidth]{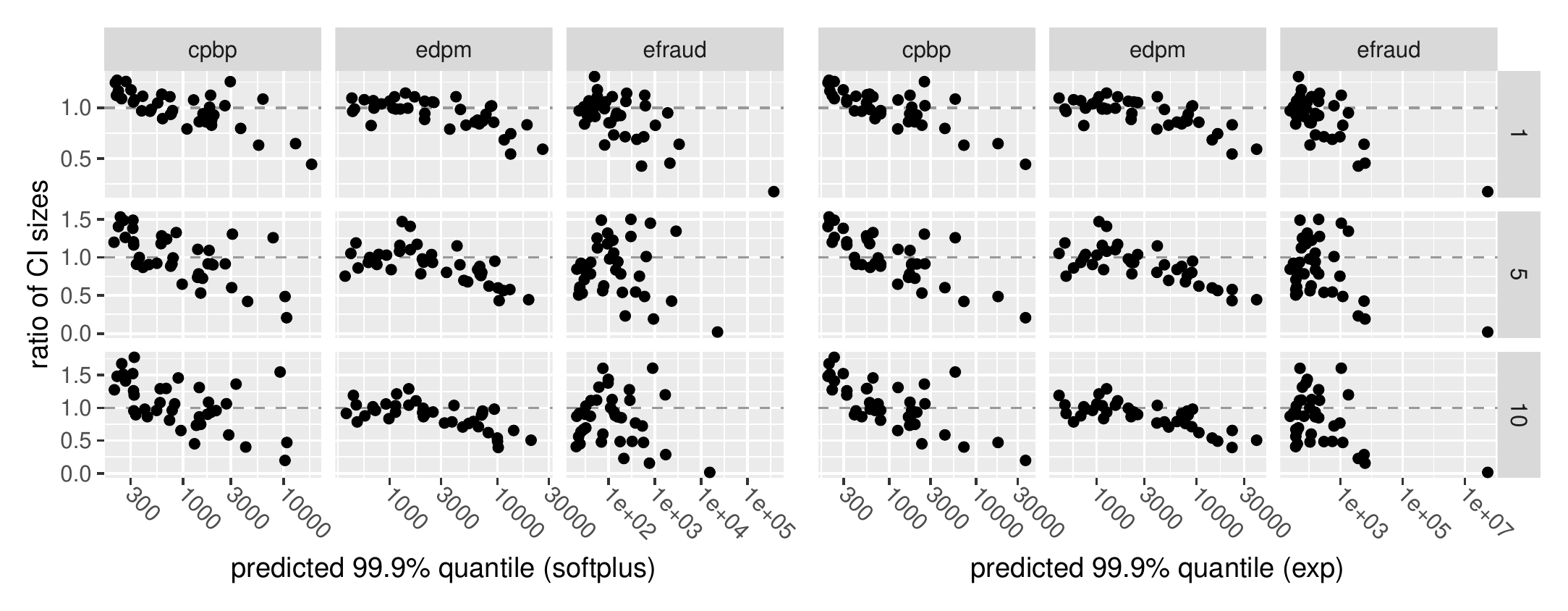}
  \caption{Ratio of the sizes of the confidence intervals (sp / exp). Left panel: ratio expressed as a function of the softplus posterior mean estimate. Right panel: ratio expressed as a function of the exponential posterior mean estimate. Each row displays the results with the softplus parameter set to the value indicated on the right ($a = 1, 5, 10$).}\label{fig:gpd_CI}
\end{figure}

However, looking first at the predicted values of the 99.9\% quantile of the conditional distributions, models based on the softplus functions generate less outliers than the exponential function (Figure~\ref{fig:gpd_boxplot}): the largest predicted quantiles are between 1.5~and 3471~times smaller with the softplus models than with the exponential model.
Whereas UniCredit is exposed to extremely high (and unrealistic) capital requirements if it uses the exponential model, this issue is well mitigated with the softplus model.
This effect is particularly strong for \texttt{EFRAUD}.
Second, looking at the size of the confidence intervals for the 99.9\% quantiles, we observe a clear trend: on Figure~\ref{fig:gpd_CI}, we show the ratio between the size of the confidence intervals obtained with the softplus functions and those obtained with the exponential function, with values smaller than 1 indicating an advantage for the softplus function.
For large values of the estimated quantile, we obtain much narrower confidence intervals with the softplus functions, with most ratios below 1.
This result implies that, in times of financial stress characterized by high values of the quantiles, the softplus models delivers more informative estimations.

\begin{figure}[tbp]
\includegraphics[width=\textwidth]{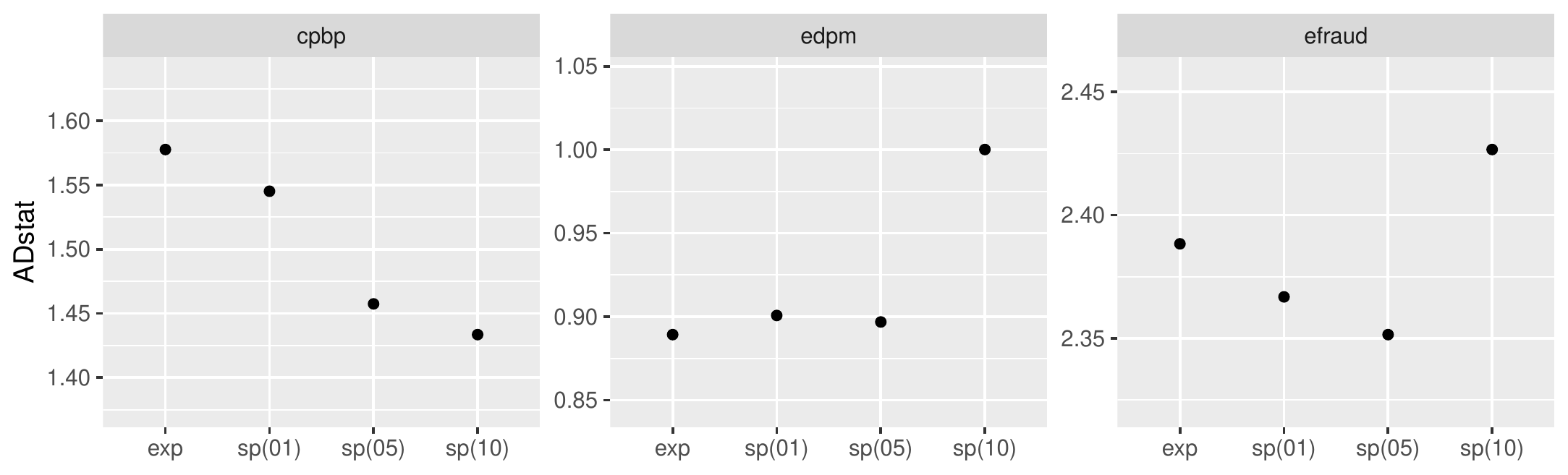}
\caption{Anderson-Darling statistics obtained from the RQRs for the three categories.}\label{fig:ad_stat}
\end{figure}

Finally, we investigate if these results imply also a better fit of the models based on the softplus function for the observations far in the tail. To do so, we report the Anderson-Darling~(AD) statistics~\citep[][]{stephensEDFStatisticsGoodness1974} computed on the RQRs (Figure \ref{fig:ad_stat}). Compared to AIC or DIC, the AD statistic gives more weight to extreme residuals and is therefore routinely used to assess the goodness-of-fit of extreme value regression models \citep{choulakian2001,bader2018}. We observe a better fit of the sofplus functions for CPBP and EFRAUD categories. The fit is rather similar for EDPM, although slightly better for the exponential model.

Overall, this application demonstrates the usefulness of the softplus function to prevent outliers among estimated quantities of interests (in the present case, a quantile far in the tail) when there are no justifications for a multiplicative model.
In addition, it shows that the softplus models provide similar global goodness-of-fit levels but dramatically reduce the uncertainty around large estimated quantities of interest, a desirable feature for end-users.

\section{Summary and Conclusion}\label{sec:sumconc}

In this paper, we introduce softplus response function and showcase its applicability in a broad range of statistical models.
The novel response function ensures the positivity of the associated distribution parameter while allowing for a quasi-additive interpretation of regression effects for the majority of the relevant predictor space.
We highlight the interesting theoretical properties of the sofplus response function, justify the quasi-additive interpretation and give a guideline to assess the validity of this interpretation.

Particular emphasis is placed on demonstrating the straightforward quasi-linear and quasi-additive interpretation of covariate effects with several applications.
Furthermore, we highlight that the limited growth rate of the softplus response function can prevent outliers in predictions and, thus, can reduce prediction uncertainty.
Thereby, we show that the new response function is applicable to a great variety of model classes and data situations.

Our simulation studies demonstrate that the softplus function behaves well as a response function with no noticeable shortcomings.
Estimates are consistent and our Bayesian approach yields reliable credible intervals.
Furthermore, we show that information criteria can be used to distinguish between data generated by the exponential and the softplus response function.

We do not claim that the softplus function is, in general, a better response function than the exponential function.
However, we feel that having a quasi-additive alternative available is useful at least for empirical verification.
Since the implementation of the softplus response function is easy and straightforward,
we are optimistic that it will be available in more software for regression modeling (it is already included in the R-packages \texttt{bmrs} \citep{JSSv080i01} and \texttt{bamlss} \citep{umlaufBAMLSSBayesianAdditive2018}).
Thus, researchers can benefit from employing the softplus response function in the future.
The work of \citet{weissSoftplusINGARCHModel2021}, evaluating the softplus response function in the context of INGARCH models, is a first indication for interest of the statistical community in the novel response function.

\section*{Acknowledgements}
Financial support from the German Research Foundation (DFG) within the research project KN 922/9-1 is gratefully acknowledged.
Julien Hambuckers acknowledges the financial support of the National Bank of Belgium.
\bibliographystyle{foo}
\bibliography{softplus,softplus_julien}
\end{document}

%% file: pkg_def.tex
\usepackage{amsmath}
\usepackage{amssymb}
\usepackage{bm}
\usepackage{dsfont}
\usepackage{enumitem}
\usepackage{authblk}
\usepackage{lipsum}
\usepackage{booktabs}

\DeclareBoldMathCommand\betavec{\beta}
\DeclareBoldMathCommand\epsvec{\varepsilon}
\DeclareBoldMathCommand\muvec{\mu}
\DeclareBoldMathCommand\thetavec{\theta}
\DeclareBoldMathCommand\zerovec{0}
\DeclareBoldMathCommand\onevec{1}
\DeclareBoldMathCommand\xvec{x}
\DeclareBoldMathCommand\yvec{y}
\DeclareBoldMathCommand\gvec{g}
\DeclareBoldMathCommand\Xmat{X}
\DeclareBoldMathCommand\Imat{I}
\DeclareBoldMathCommand\Fmat{F}

\DeclareMathOperator{\softplus}{softplus}

\DeclareMathOperator{\rect}{rect}
\DeclareMathOperator{\h}{h}



%% file: abstract.tex
Response functions linking regression predictors to properties of the response distribution are fundamental components in many statistical models.
However, the choice of these functions is typically based on the domain of the modeled quantities and is not further scrutinized.
For example, the exponential response function is usually assumed for parameters restricted to be positive although it implies a multiplicative model which may not necessarily be desired.
Consequently, applied researchers might easily face misleading results when relying on defaults without further investigation.
As an alternative to the exponential response function, we propose the use of the softplus function to construct alternative link functions for parameters restricted to be positive.
As a major advantage, we can construct differentiable link functions corresponding closely to the identity function for positive values of the regression predictor, which implies an quasi-additive model and thus allows for an additive interpretation of the estimated effects by practitioners.
We demonstrate the applicability of the softplus response function using both simulations and real data.
In four applications featuring count data regression and Bayesian distributional regression, we contrast our approach to the commonly used exponential response function.

%% file: arxiv-files/hs_dic_tab.tex
\begin{table}[tb]
\centering
\begin{tabular}{lrrr}
  \toprule
 & negbin (ZA) & negbin\\
  \midrule
  exp & 716 & 740\\
  softplus & \underline{715} & 738\\
   \bottomrule
\end{tabular}
\caption{The table displays the DIC values broken down by response function and response distribution for each model fitted. ZA indicates the zero-adjusted response distribution.} 
\label{tab:hs-dic}
\end{table}

%% file: arxiv-files/hs_coef_tab.tex
\begin{table}[b]
\centering
\begin{tabular}{r|rrr|rrrr}
  \toprule
  & \multicolumn{3}{c|}{softplus} & \multicolumn{4}{c}{exponential}\\
  & Mean & 2.5\% & 97.5\% & Mean & 2.5\% & 97.5\% & $\exp(\cdot)$ \\
  \midrule
  (Intercept) & -9.65 & -15.86 & -3.59 & -3.12 & -5.66 & -0.86 & 0.08 \\
  width & 0.53 & 0.32 & 0.75 & 0.18 & 0.11 & 0.27 & 1.20 \\
  color & -0.54 & -1.12 & 0.01 & -0.27 & -0.52 & -0.04 & 0.77 \\
  $\log(\sigma)$ & 1.13 & 0.74 & 1.53 & 1.14 & 0.76 & 1.51 &  - \\
   \bottomrule
\end{tabular}
\caption{Posterior estimates of the regression coefficients on the expected value together with their 95\% credible intervals (equal-tailed). The last column shows the posterior mean of the exponential function applied to the regression coefficient. Besides, $\sigma$ denotes the dispersion parameter.} 
\label{tab:hs-coef}
\end{table}

%% file: arxiv-files/cbs-sigma-le.tex
\begin{table}[bth]
\centering
\begin{tabular}{l|rrrr|rrr}
    \toprule
    & \multicolumn{4}{c|}{exponential} & \multicolumn{3}{c}{softplus} \\
   & Mean & 2.5\% & 97.5\% & $\exp(\cdot)$ & Mean & 2.5\% & 97.5\%\\
    \midrule
(Intercept) & 0.151 & 0.134 & 0.168 & 1.163 & 1.325 & 1.303 & 1.348\\ 
  weekend & 0.392 & 0.364 & 0.419 & 1.480 & 0.566 & 0.525 & 0.604\\ 
  year2017 & -0.016 & -0.037 & 0.005 & 0.984 & -0.009 & -0.032 & 0.014\\ 
   \bottomrule
\end{tabular}
\caption{Posterior estimates of the linear effects on the predictor of the standard deviation together with their 95\% credible intervals (equal-tailed).} 
\label{tab:cbs-sigma-le}
\end{table}

%% file: arxiv-files/ol_desc_stats.tex
\begin{table}[htb]
\centering
\begin{tabular}{l|cccccc}
  \toprule
Category & n & mean & median & skewness & kurtosis & iqr \\ 
  \midrule
  \texttt{CPBP} & 4034 & 255879 & 29453 & 24 & 674 & 68361 \\ 
  \texttt{EDPM} & 3302 & 133468 & 15571 & 20 & 539 & 43689 \\ 
  \texttt{EFRAUD} & 1598 & 64027 & 15277 & 37 & 1412 & 31169 \\ 
  \bottomrule
\end{tabular}
\caption{Descriptive statistics of the exceedance. \emph{iqr} denotes the inter-quantile range.} 
\label{tab:ol-desc}
\end{table}